\documentclass[preprint,12pt,authoryear]{elsarticle}

\usepackage{amssymb}
\usepackage{amsmath}
\usepackage{subfig}
\usepackage{algpseudocode}
\usepackage[lined,boxed,ruled,commentsnumbered]{algorithm2e}
\newtheorem{theorem}{Theorem}

\begin{document}

\begin{frontmatter}



\title{RONAALP: Reduced-Order Nonlinear Approximation with Active Learning Procedure}


\author[IJLRA]{Cl\'ement Scherding}
\ead{clement.scherding@dalembert.upmc.fr}
\author[ICL]{Georgios Rigas}
\author[ONERA]{Denis Sipp}
\author[KAUST]{Peter J. Schmid}
\author[IJLRA,ITW]{Taraneh Sayadi}

\affiliation[IJLRA]{organization={Institut Jean le Rond d'Alembert, Sorbonne University},
            addressline={}, 
            city={Paris},
            postcode={75005}, 
            state={},
            country={France}}

\affiliation[KAUST]{organization={Department of Mechanical Engineering, KAUST},
            addressline={}, 
            city={Thuwal},
            postcode={23955}, 
            state={},
            country={Saudi Arabia}}

\affiliation[ICL]{organization={Department of Aeronautics, Imperial College},
            addressline={}, 
            city={London},
            postcode={SW7 2AZ}, 
            state={},
            country={United Kingdom}}

\affiliation[ONERA]{organization={DAAA, Onera},
            addressline={}, 
            city={Meudon},
            postcode={92190}, 
            state={},
            country={France}}

\affiliation[ITW]{organization={Institute for Combustion Technology, Aachen University},
            addressline={}, 
            city={Aachen},
            postcode={52062}, 
            state={},
            country={Germany}}        

\begin{abstract}

Many engineering applications rely on the evaluation of expensive, non-linear high-dimensional functions. In this paper, we propose the RONAALP algorithm (\textbf{R}educed \textbf{O}rder \textbf{N}onlinear \textbf{A}pproximation with \textbf{A}ctive \textbf{L}earning \textbf{P}rocedure) to incrementally learn a fast and accurate reduced-order surrogate model of a target function on-the-fly as the application progresses. First, the combination of nonlinear auto-encoder, community clustering and radial basis function networks allows to learn an efficient and compact surrogate model with limited training data. Secondly, the active learning procedure overcome any extrapolation issue when evaluating the surrogate model outside of its initial training range during the online stage. This results in generalizable, fast and accurate reduced-order models of high-dimensional functions. The method is demonstrated on three direct numerical simulations of hypersonic flows in chemical nonequilibrium. Accurate simulations of these flows rely on detailed thermochemical gas models that dramatically increase the cost of such calculations. Using RONAALP to learn a reduced-order thermodynamic model surrogate on-the-fly, the cost of such simulation was reduced by up to 75\% while maintaining an error of less than 10\% on relevant quantities of interest. 

\end{abstract}




\end{frontmatter}


\section{Introduction}
\label{intro}

In various engineering applications that involve numerical simulations, there is a frequent need to assess a function, denoted as $f(\textbf{x})$, numerous times. For instance, when solving partial differential equations with a given numerical discretization (eg. finite difference), it becomes necessary to evaluate the function $f$ for every grid node at each time step, considering the current values of dependent variables ($\textbf{x}$) at those nodes. If evaluating the function $f$ is computationally intensive, these function calls become the main bottleneck of the program. Therefore, it is logical to explore more cost-effective methods for evaluating $f$ and obtaining an approximation that is sufficiently accurate. This extremely general problem is known as Reduced-Order Modeling (ROM), or surrogate modeling, of the Full-Order Model (FOM), $f$.

One instance of a problem where a ROM would be beneficial in reducing the computational cost is the numerical simulation of hypersonic flows. In fact, an object flying at hypersonic speed is surrounded by an extremely complex flow environment. The dissipation of kinetic energy introduces highly energetic gas states. These high-temperature states evolve in a flow dominated by extremely short time scales. At the molecular level, there is no guarantee that the collisions are sufficiently frequent for the energy exchange and chemical process to reach equilibrium. Thus, the composition and properties of the gas can vary in space and time, and the flow departs from thermal and/or chemical equilibrium. These nonequilibrium effects can significantly influence the flow behavior, heat transfer, and chemical kinetics in hypersonic environments \citep{Holden1986,Leyva2017}. In recent years, there has been increasing interest in understanding the nature and impact of thermochemical nonequilibrium in hypersonic flows \citep{johnson1998numerical,Marxen2013,direnzo2021,passiatore2022}, leading to the comprehensive review of \cite{Candler2019}. The numerical simulation of the complex interplay between thermodynamics, chemistry and fluid mechanics in these extreme conditions is a difficult modeling challenge and relies on detailed thermochemical gas models that dramatically increase the cost of such calculations. The thermodynamic library Mutation++ \citep{scoggins2020} was used in the present work as the reference high-fidelity model $f$.

Relatively few studies have tried to make simulations in that flow regime faster. For example, \cite{mao2021deepm} and \cite{gkimisis2023data} used respectively DeepONet and artificial neural network (ANN) to predict the coupled flow in chemical nonequilibrium past a normal shock.  \cite{zanardi2022towards} used physics-informed DeepONet to reduce the stiff master equations (equivalent to the thermochemical model in a state-to-state kinetic framework) into a ML-based surrogate. These authors reported up to 2 orders of magnitude faster prediction time but these studies were restricted to simple 0D or 1D configurations. Alternatively, \cite{scherding2023data} have recently proposed a data-driven framework for extracting a fast and accurate reduced-order thermochemical gas model. The lightweight model is constructed using thermochemical states obtained from a direct numerical simulation (DNS) in the parameter regime considered. These states are embedded in a low-dimensional subspace through the use of a deep encoder and subsequently clustered into regions at different levels of thermochemical (non)equilibrium. Finally, several surrogate models are constructed for each cluster in the low-dimensional subspace. The method was validated on two-dimensional laminar simulations of a Mach $10$ adiabatic boundary layer and Mach $5.92$ shock wave boundary layer interaction in chemical nonequilibrium (CNEQ). The combination of the preprocessing steps enabled the construction of faster and highly accurate reduced-order thermochemical model. The models were trained on the converged baseflow solution and were shown to maintain a stable solution after restarting the simulation with the reduced-order thermochemical model, while reducing by up to 70\% the evaluation time of thermochemical properties. 

Hypersonic flows can exhibit unsteady features such as hydrodynamic instabilities and turbulence. Turbulence enhances the average skin friction and heat-flux at the wall -- two critical design parameters -- compared to a laminar flow. High-order numerical methods are well suited to study such flows with their high accuracy and minimal modeling assumptions. However, the stringent requirement for performing direct numerical simulations of turbulent hypersonic flows and the additional cost incurred by the complex thermochemical nonequilibrium model have limited numerical studies to the lower Reynolds number regime and simple configurations \citep{direnzo2021,passiatore2022}. Therefore, an optimized and tailored thermochemical model could be utilized to study higher Reynolds number flows, for instance. 

However, unsteadiness modifies the thermodynamic manifold. A model trained on a steady solution will likely extrapolate beyond its training range due to the presence of new thermodynamic states pertaining to the flow unsteady characteristics, as shown in \cite{scherding2022ctr}, making direct application of the developed strategy not straight forward. 
Even with thorough training, ROMs typically face limitations when it comes to extrapolation beyond their training range. This issue is widely recognized as a common challenge for data-driven techniques. Consequently, ROMs cannot be considered predictive or generalizable since their outputs may deviate significantly and yield incorrect results when confronted with inputs or conditions that fall outside the range of observed data. Expanding the training range with additional realizations is often too expensive, particularly when it involves cost-intensive CFD simulations. In \cite{mao2021deepm,zanardi2022towards,gkimisis2023data}, the authors used inexpensive 0D and 1D to bloat the training range. This strategy is however impractical for the 2D and 3D DNS tackled in \cite{scherding2023data} and would overall counteract the purpose of the ROM in speeding up simulations. 

To mitigate this issue, additional techniques and approaches have been proposed to enhance the ROMs' extrapolation capabilities. To that end, adaptive reduced-order models are a promising solution. Rather than being confined to a specific operating window or application, they have the capacity to learn dynamically and refine the model on-the-fly, enabling broader applicability and superior results in a wide range of regimes. In the dynamical systems community, recent research focuses on building adaptive ROMs that can learn new dynamics in situ \citep{peherstorfer2020, yano2021, ramezanian2021, huang2023predictive}. The ISAT algorithm, \cite{pope1997computationally}, is an adaptive look-up table method that can learn in situ new inputs/outputs relation if an error metric is satisfied. However, the algorithm relies on linear interpolation only. Radial basis functions networks, a type of universal approximator, capable of on-the-fly learning have also seen some development \citep{platt1991, kadirkamanathan1993function, karayiannis1997growing, huang2005generalized,bortman2009growing}.

The goal of this work is therefore to enhance the data-driven framework presented in \cite{scherding2023data} through the integration of an active learning procedure, ensuring the generalizability and predictive capabilities of the resulting model. This lead to the development of the RONAALP algorithm (acronym for \textbf{R}educed \textbf{O}rder \textbf{N}onlinear \textbf{A}pproximation with \textbf{A}ctive \textbf{L}earning \textbf{P}rocedure). We believe that the algorithm can be readily ported and benefit other fields relying on expensive high-dimensional functions evaluation. 

The paper is organized as follows. In Section \ref{goveq}, the governing equations and thermochemical modeling of hypersonic flow in chemical nonequilibrium are recalled. The off-line construction of the data-driven model is briefly presented in Section \ref{algo} before presenting the active learning procedure. Finally, this novel method is tested in Section \ref{tests} for three time marching simulations that generate thermodynamic states unseen during training. Namely, a low-fidelity (self-similar) to high-fidelity DNS transient simulation and an optimally disturbed laminar boundary layer, both in a 2D and 3D set-up, initially studied by \cite{Marxen2013,Marxen2014a}. The resulting flow dynamics are then compared to the baseline simulation using Mutation++. Finally, conclusions are drawn in Section \ref{conclusion}. 

\section{Modeling of hypersonic flows in chemical nonequilibrium}
\label{goveq}

In the following, we define the governing equations and the relevant thermochemical model for the simulation of hypersonic flows in chemical nonequilibrium. The numerical framework used to perform direct numerical simulation (DNS) of such flows is also described.

\subsection{Governing equations}
We consider the reactive compressible Navier-Stokes equations for an air mixture of five species $S = \{$N, O, NO, N$_2$, O$_2\}$ given as 

\begin{equation}\label{continuity}
\frac{\partial \rho}{\partial t} + \mathbf{\nabla}\cdot(\rho \mathbf{u}) = 0,
\end{equation}
\begin{equation}\label{species}
\frac{\partial \rho_s}{\partial t} +  \mathbf{\nabla}\cdot(\rho_s (\mathbf{u}+\mathbf{V}_s)) = \dot{\omega}_s, \quad s \in S, 
\end{equation}
\begin{equation}\label{momentum}
\frac{\partial \rho \mathbf{u}}{\partial t} + \mathbf{\nabla} \cdot (\rho \mathbf{u} \otimes \mathbf{u} )= -\mathbf{\nabla} P + \mathbf{\nabla}\mathbf{\tau},
\end{equation}
\begin{equation}\label{energycons}
\frac{\partial \rho e_0}{\partial t} + \mathbf{\nabla} \cdot (\rho (e_0 + P)\mathbf{u}) =  \mathbf{\nabla} \cdot (\mathbf{\tau} \cdot \mathbf{u}) - \mathbf{\nabla} \cdot \mathbf{q},
\end{equation}
where the corresponding velocity components are $\mathbf{u}=\{u,v,w\}$, $t$ denotes time, $\rho$ denotes the mixture density, and $\rho_s = \rho Y_s$ and $Y_s$ are the partial density and mass fraction of species $s \in S$, respectively. These equations are integrated numerically for all but one species $s \in S$ in a three-dimensional Cartesian coordinate system $\{x,y,z\}$. $x,y,$ and $z$ point in the streamwise, wall-normal, and spanwise directions, respectively. In the momentum equation (Eq.~(\ref{momentum})), $P$ stands for pressure, and $\mathbf{\tau}$ is the viscous stress tensor, defined for a Newtonian fluid as
\begin{equation}
    \mathbf{\tau} = \mu (\mathbf{\nabla}\mathbf{u} + \mathbf{\nabla}\mathbf{u}^T - 2(\nabla \cdot \mathbf{u})\mathbf{I}/3),
\end{equation}
where $\mu$ is the dynamic viscosity and $\mathbf{I}$ is the identity tensor. In the energy equation (Eq.~(\ref{energycons})), $e_0 = e + \lvert \mathbf{u} \rvert^2/2$ is the stagnation internal energy, with $e$ denoting the specific internal energy, defined from the species-specific enthalpies $h_s$ as 
\begin{equation}\label{energy}
    e = \sum_{s \in S} h_s Y_s - P/\rho.
\end{equation}
The heat flux vector takes the form
\begin{equation}\label{heatflux}
    \mathbf{q} = - \kappa \nabla T + \sum_{s\in S} \rho_s h_s \mathbf{V}_s,
\end{equation}
where $T$ denotes the temperature. The term $\mathbf{V}_s$, appearing in Eqs.~(\ref{species}) and (\ref{heatflux}), denotes the diffusion velocity vector of species $s$. The diffusion velocities can be rigorously calculated as the solution of a constrained linear system of equations, known as the Stefan-Maxwell multicomponent diffusion model \cite{Scoggins2017}. However, this model includes local molar fraction gradients which cancels the purely local input-output assumption of the library. Hence, in this study it is modeled as a Fickian flux with a mass correction term defined as \citep{hirschfelder1964,ramshaw1990}
\begin{equation}\label{ramshaw}
   \rho_s \mathbf{V}_s = -c W_s D_s \nabla Y_s + c Y_s \sum_{i \in S} W_i D_i \nabla Y_i .
\end{equation}
Here, $c=\sum_{s\in S} \rho_s / W_s$, where $W_s$ is the individual species molecular weight. $D_s$ is the averaged diffusion coefficient for species $s$ based on individual binary diffusivities $D_{s,i}$, $~\{s,i\} \in S$, computed following the rule proposed in \cite{hirschfelder1964} as,
\begin{equation}
   D_s = \frac{1-X_s}{\sum_{i \neq s} X_i/D_{s,i}},
\end{equation}
where $X_s=Y_s \overline{W}/W_s$ is the mole fraction of species $s$ and $\overline{W} = (\sum_{s \in S} Y_s / W_s)^{-1}$ the mixture averaged molecular weight. This comparatively simpler diffusion model has shown to be accurate for hypersonic flows considered in this paper \citep{margaritis2022}.
Finally, the net mass production rate of species $\dot{\omega}_s$ in Eq.~(\ref{species}), considering all reactions, is computed using \cite{park1989} five-reactions chemical mechanism for dissociated air. 

\subsection{Thermochemical model}
The governing equations are closed using the equation of state
\begin{equation}
    P = \rho R_u T / \overline{W},
   \label{eq:state}
\end{equation}
where $R_u$ is the universal gas constant. The transport ($\mu, \kappa, D_s$), thermodynamic ($P,T,h_s$), and chemical $\dot{\omega}_s$ properties defined above are generally a function of two independent thermodynamic state variables and the mixture composition, and need to be modeled accordingly. Different modeling approaches exist in the literature \citep{direnzo2021,passiatore2022}, with the main drawback being that the thermochemical model has to be hard-coded. Hence, any update in the model, for instance, to add complexity or to simulate a different mixture, comes at a human cost in terms of implementation, testing, and validation. These limitations led to the development of the library Mutation++ \citep{scoggins2020}, which offers a flexible high-level application programming interface to model the physico-chemical properties of mixtures in different levels of non-equilibrium. A wide range of algorithms for the calculation of the individual and mixture-averaged properties are supported. The library is easily coupled to any computational fluid dynamics solver as an input/output problem $\mathbf{z} = f(\mathbf{x)}$. More precisely, given the local state vector
\begin{equation}
    \mathbf{x} = [\rho, \rho_s, \rho e] \in \mathbb{R}^D, 
\end{equation}
the library returns all physico-chemical properties needed to close the governing equations 
\begin{equation}
    \mathbf{z} = f(\mathbf{x}) = [P, T, \mu, \kappa, D_s, h_s, \dot{\omega}_s] \in \mathbb{R}^{D_Z}. 
\end{equation}

\subsection{Numerical framework}
The compressible reactive Navier-Stokes equations are solved using a high-order finite-difference method together with a fourth-order explicit Runge-Kutta time integration on staggered grids. The solver is coupled with the Mutation++ library for simulations in chemical non-equilibrium but can also run with calorically or thermally perfect gas assumptions. Several test cases of canonical hypersonic flows, including the one used in Section \ref{tests}, are presented and validated in \cite{margaritis2022}. 
The solution are initialized with locally self-similar solutions in chemical non-equilibrium \citep{lees1956,williamsCTR2021}. While close to the actual solution of the Navier-Stokes equations, they miss some physics, such as streamwise species diffusion. This leads to a transient stage of the numerical solution from the self-similar towards the true Navier-Stokes solutions. However, self-similar solutions still provide a good approximation of the thermodynamic manifold of the true solution, as shown in~\cite{scherding2022ctr}. Therefore, they can be used to warm-start the off-line training of the model.

\section{RONAALP -- algorithm}
\label{algo}
\begin{figure}
    \centering
    \includegraphics[width=0.6\textwidth]{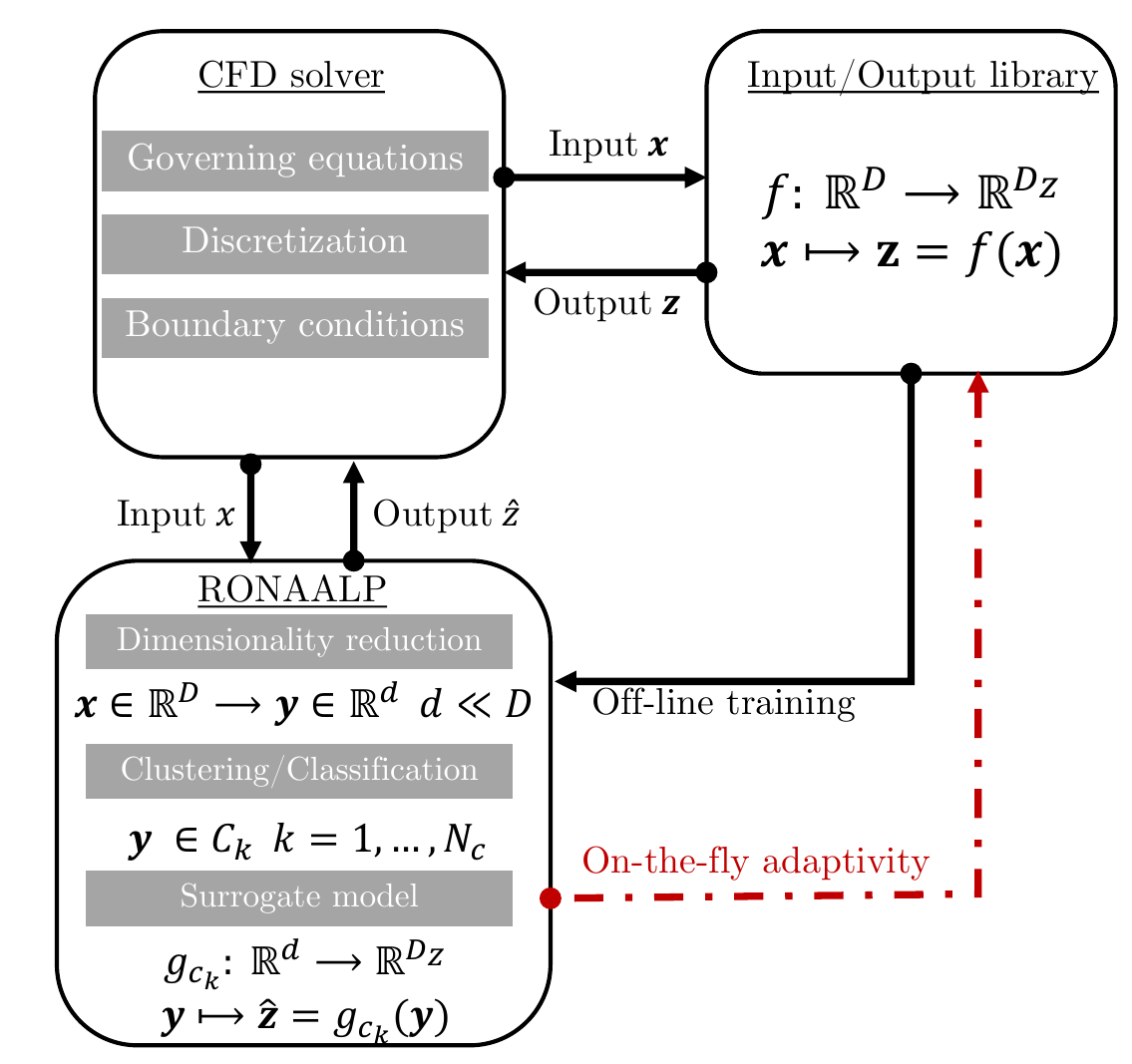}
    \caption[General schematic of the model training and coupling with adaptivity.]{General schematic of the model training and coupling to replace any expensive Input/Output library.} 
    \label{fig:algo_adapt}
\end{figure}

In this section, we first recall the off-line learning strategy, thoroughly described in \cite{scherding2023data}. Secondly, we define the methodology to enable active learning capability of the reduced-order model, which describes the parts highlighted in red in the schematic of the data-driven model in Figure \ref{fig:algo_adapt}. 

To showcase the different steps, the off-line training uses locally self-similar solutions (low-fidelity data) while the active learning procedure is showcased on data from a converged laminar DNS solution (high-fidelity data).

\subsection{Off-line training}
\label{offline}
\begin{figure}
\begin{center}
    \includegraphics[width=1.1\textwidth]{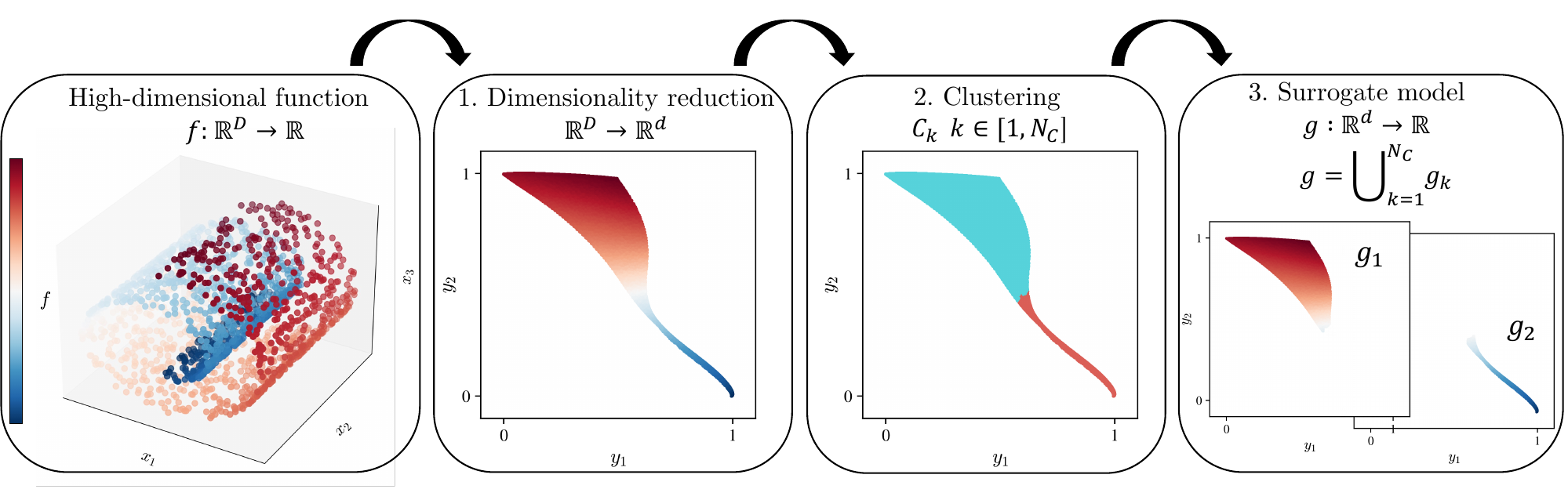}
        \caption[]{General schematic of the model training to replace any expensive input/output library.}
        \label{sketch}
	\end{center}
\end{figure}

Mathematically, the data-driven model $g$ should predict the outputs of the function of interest $\hat{\mathbf{z}}=g(\mathbf{x})$ such that $\lVert \hat{\mathbf{z}} - \mathbf{z} \rVert_2$ is minimized (preserving accuracy) and the computational cost is significantly reduced compared to the original library $f$. The strategy employed to derive the reduced-order model is thoroughly presented in \cite{scherding2023data} and the building of the reduced-order model is sketched in Figure \ref{sketch} for a generic high-dimensional function. 

\noindent In summary, the training is done in three sequential steps described below:
\begin{enumerate}
\item {\bf Dimensionality reduction}: For a given engineering problem, only a subset of all possible inputs is encountered. Hence, only a small subset of the library is accessed during a simulation. Moreover, the governing equations of the physical system induce a correlation between the different input variables. Theoretically, fewer variables are therefore needed to obtain full-state information. Thus, the local state vectors $\mathbf{x}$ are projected onto a low-dimensional space that preserves the variation of the outputs through an input-output encoder (IO-E). 
\begin{equation}
 E \colon \biggl\{\begin{array}{@{}r@{\;}l@{}}
   ~ \mathbb{R}^D &\to \mathbb{R}^d
  ,\\
   ~ x ~~ &\mapsto E(x).
  \end{array}
\end{equation}
The IO-E consists of two sequential deep neural networks. The first network, the encoder $E$, projects the inputs of the library in a latent space of dimension $d<D$. The second network, the decoder, predicts the outputs of the library from this latent space. The training is done through back-propagation of the $L_2$ norm of the error $\lVert \Tilde{\mathbf{z}} - \mathbf{z} \rVert_2$ through the full IO-E, where $\Tilde{\mathbf{z}}$ denotes the prediction of the network. However, only the encoder part of the network is used. In fact, the radial basis function networks (RBF) described later have shown better accuracy than the decoder on the case considered. This first pre-processing step avoids interpolation and prediction in high dimensions, a tedious task due to the curse of dimensionality. 
\item {\bf Clustering \& classification}: In the latent space, spectral clustering using \cite{newman2006} algorithm allows the determination of $N_C$ clusters, representing regions with different dynamics of the function $f$. This second pre-processing step allows for a tailored fitting over a given region, especially in the presence of discontinuities due to shocks in a hypersonic flow in CNEQ, for instance. 
\item {\bf Surrogate model}:  Finally, RBFs are constructed with $N_{R}$ centers on each spectral cluster. The radial basis function interpolant $g$ is given by
\begin{equation}
\label{eq:rbf}
   g(\phi,\textbf{x}) = \sum_{i=1}^{N_\textrm{R}} \lambda_i \phi(\lVert \textbf{x} - \textbf{x}_i^c \rVert),
\end{equation}
where $\phi$ is the kernel function, such as a Gaussian $\phi(r)=exp(-r^2/2l^2)$ or thin-plate spline $\phi(r)=r^2 log(r)$. The set of centers is denoted as $\textbf{X}^C=\{\textbf{x}_1^c,\dots, \textbf{x}_{N_R}^c \}$ and is determined via $k$-means clustering on each spectral cluster. The resulting $k$-means centroids become the centers for each RBF. The weights $\boldsymbol\Lambda = [\lambda_{1},..., \lambda_{N_\textrm{R}}]^T$ that minimize the mean-square error of the RBF over the training input points can be obtained through the solution of the linear system 
\begin{equation}
	\mathbf{\Phi} \mathbf{\Lambda} = \mathbf{f},
\end{equation}
where $\textbf{f} = [f(\textbf{x}_1^c), ..., f(\textbf{x}_{N_{R}^c})]^T$ denotes the vector containing the function values at the RBF center. The kernel matrix $\mathbf{\Phi}$ is defined as
\begin{equation}
	\mathbf{\Phi}_{i,j} = \phi(\lVert \mathbf{x}_i - \mathbf{x}_j \rVert).
\end{equation}
\end{enumerate}

\subsection{Online learning}
\label{online}

In this section, we describe the methodology for on-the-fly active learning of the model during a time-marching simulation. The strategy is divided in three steps:

\begin{enumerate}
\item \textbf{Extrapolation detection}: We first define a metric to detect when the model is extrapolating.
\item \textbf{Online k-means}: Secondly, new centers are added to the RBF using an online clustering procedure.
\item \textbf{Updating the RBF surrogate}:  Finally, the RBF is efficiently updated by taking into account the new centers. 
\end{enumerate} 

During a numerical simulation, the user sets an updating frequency $f_\textrm{up}$ (i.e. a number of iterations) at which all the steps described below are performed and the model is updated. 

\subsubsection{Extrapolation detection}
\label{sub:extrapolation_detection}
\begin{figure}
\begin{center}
    \subfloat[]{%
      \includegraphics[width=0.8\columnwidth]{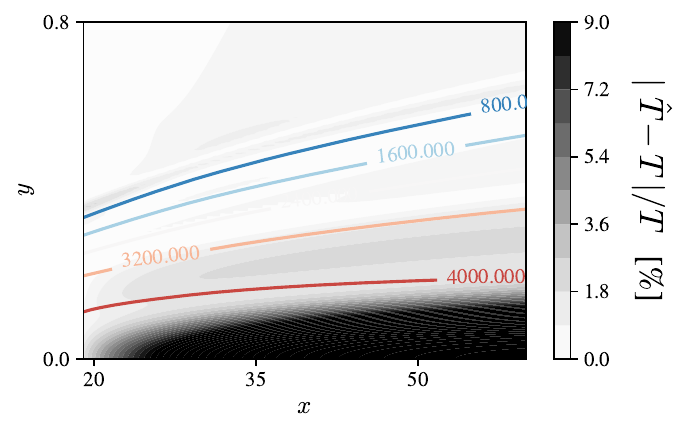}
    }\\
    \subfloat[]{%
      \includegraphics[width=0.45\textwidth]{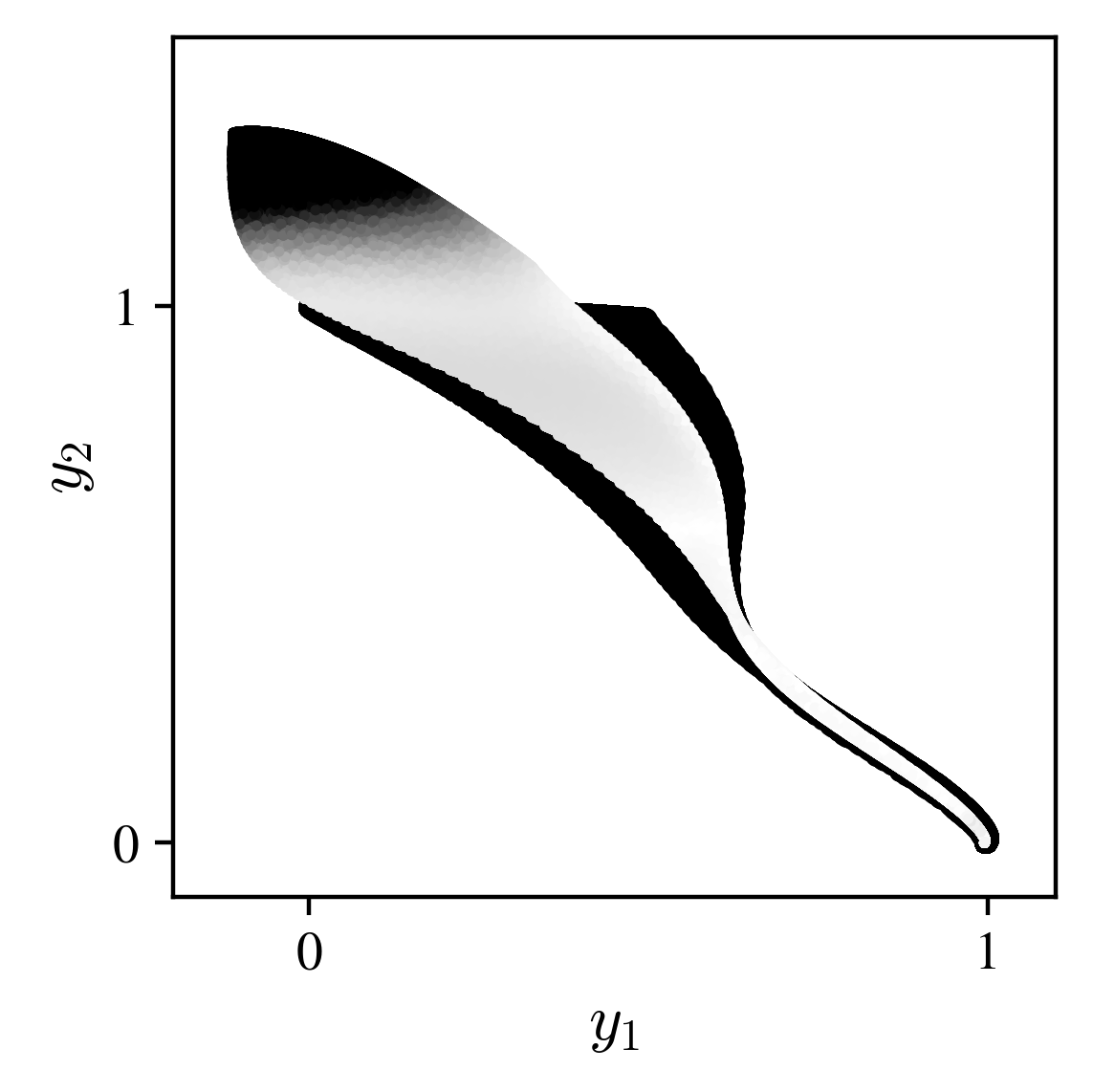}%
    }
    \caption[Error before retrain.]{Relative temperature error $\lvert \hat{T}-T \rvert/T$ in percent with contours of temperature before retraining the model: (a) the physical space of the simulation; (b) latent space with the training data shown in the background in plain black.}\label{fig:error_before_retrain}
\end{center}
\end{figure}

To overcome the generalization problem, one should first describe the region of the embedded space where training data is available. In fact, a correct characterization of this region would allow us to identify the areas where the outputs of the RBF are not reliable, which would subsequently signal the need for retraining. Figure \ref{fig:error_before_retrain}(a) shows the relative error field of the data-driven model when predicting temperature on the converged DNS solution (true solution) while the model was trained on the self-similar solution only. Near the wall, high errors are observed due to the missing physics of the low-fidelity self-similar solution used for training. In the low-dimensional latent space, shown in Figure \ref{fig:error_before_retrain}(b), the high error region corresponds to a region lacking training data, shown in the background using plain black. 

One way to detect the borders of the space spanned by the training data is to use an $\alpha$-shape of this set. Briefly, an $\alpha$-shape is a generalization of the convex-hull of a set of points and was introduced by \cite{edelsbrunner1983shape}. However, the hyperparameter $\alpha$ is difficult to tune, and the method lacks generalization in higher dimensions. \cite{leonard1992neural} instead estimated the local density of training data using kernel density estimation (KDE). A low density of probability indicates possible extrapolation. An even simpler method to detect extrapolation online was a procedure proposed by \cite{lohninger1993evaluation}, which is based on RBF with gaussian kernels. Since  a function is evaluated using the distance of the evaluation point $\mathbf{x}^t$ to the centers $\mathbf{X}^c$ of the RBF (Eq.~(\ref{eq:rbf})), the difference of the maximum of the activation functions to $1$ can then be used as a parameter to flag extrapolation
\begin{equation}
	f_e(\mathbf{x}^t) = 1 - \max_{\mathbf{x}^c ~\in ~\mathbf{X}^c } \phi( \lVert \mathbf{x}^t - \mathbf{x}^c \rVert).
	\label{eq:extrapolation_flag}
\end{equation}
Using the above relation, if the evaluation point is geometrically close to a center, the second term on the right-hand side tends to $1$ and $f_e$ tends to $0$. However, if the evaluation point lies far from any center, $f_e$ tends to $1$, indicating extrapolation. A threshold on the value of $f_e$ is then used to distinguish between extrapolation and interpolation regions, respectively. A drawback of this method is that it relies on monotonic kernel functions, which is not always the case, for example when using the thin-plate spline kernels. 

Hence, we propose a more general approach based on the minimum distance of the evaluation point $\mathbf{x}^t$ to the set of centers $\mathbf{X}^c$,
\begin{equation}
	f_e(\mathbf{x}^t) = \min_{ \mathbf{x}^c ~\in~\mathbf{X}^c }  \lVert \mathbf{x}^t - \mathbf{x}^c \rVert.
\end{equation}
The extrapolation threshold then becomes dependent of the geometry of the euclidean space where the low-dimensional manifold lies. Let $\mathbf{X}^c_{i,k} = [\mathbf{x}^c_{i,1} \dots \mathbf{x}^c_{i,k}]$ be the matrix containing the $k$-nearest neighbors of centroid $\mathbf{x}^c_i$ in $\mathbf{X}^c$. The threshold for extrapolation detection, $d_e$, is then computed as 
\begin{equation}
	d_e = \frac{1}{N_\textrm{R}}\sum_{i=1}^{N_\textrm{R}}  \left( \frac{1}{k}\sum_{j=1}^{k} \rVert \mathbf{x}^c_i - \mathbf{x}^c_{i,j} \lVert \right)
\end{equation}

\begin{figure}[htbp]
    \centering
    \includegraphics[width=0.45\textwidth]{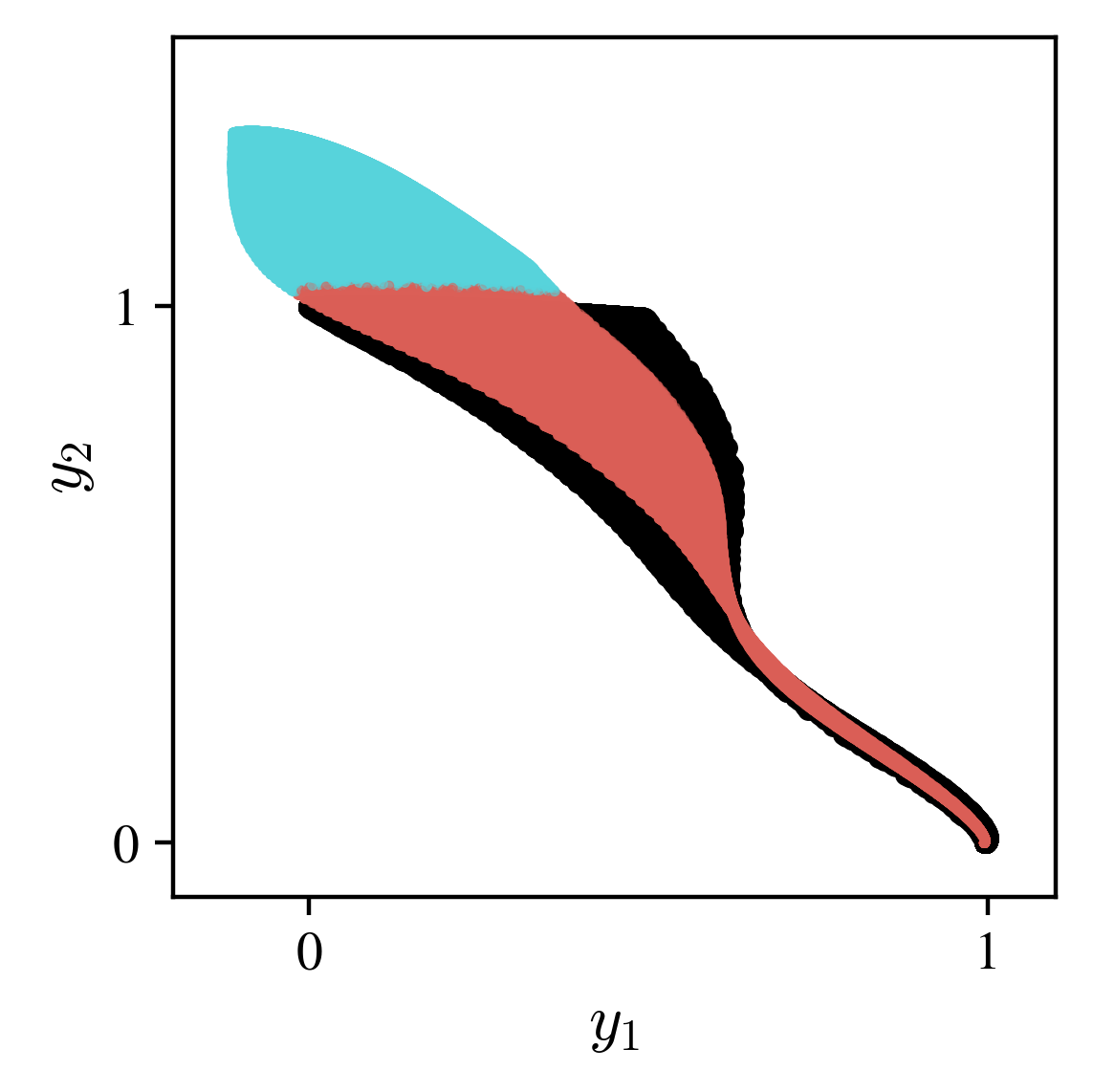}
    \caption[Detection of extrapolation regions.]{Evaluation points $\mathbf{Y}^t$ colored by the extrapolation flag: blue and red represent points detected outside and inside the training range, respectively. The $\alpha$-shape of the training data is drawn underneath in black.}\label{fig:extrapolation}
\end{figure}

A demonstration of this method is plotted in Figure \ref{fig:extrapolation}, where the points outside of the $\alpha$-shape of the training data (shown in black) are correctly flagged as extrapolation.

\subsubsection{Growing RBF}
The input/output pairs detected in extrapolation have to be learned by the RBF to improve the mapping in the newly explored region of the latent space. This topic is known as growing RBF, or more generally as resource-allocating network, and several training techniques have been proposed in the literature for their on-line training \citep{platt1991, kadirkamanathan1993function, karayiannis1997growing, huang2005generalized,bortman2009growing}. The common strategy is to sequentially feed new observations to the network. If an observation makes a significant contribution to the overall performance of the model (it fulfills the novelty criterion given a certain metric), then a new center (or neuron) is added in the hidden layer. If not, the parameters of the network in the vicinity of the observation are updated. Recent improvements even allow to prune the network, providing an optimal architecture in terms of complexity (number of units in the hidden layer) \citep{huang2005generalized,bortman2009growing}. 

The original implementation used least mean squares filters (LMS) \citep{platt1991} for the update procedure of the network parameters. While the LMS algorithm iteratively updates the filter coefficients efficiently, it does not guarantee optimality in the least-square sense. In fact, it is a stochastic gradient descent algorithm and can therefore converge to a local minimum or exhibit some residual error even after convergence. In contrast, the optimal solution in the least-square sense can be obtained using methods like the Recursive Least Squares algorithm. This method provides a closed-form optimal solution in the least-square sense. However, they are computationally expensive (due to the necessity of performing a matrix inversion) and may not be suitable for real-time applications. 
\cite{kadirkamanathan1993function} and \cite{bortman2009growing} instead used extended Kalman filters for the update procedure. This provided a good trade-off between computational efficiency and optimality of the solution (even though not strictly) in the least-square sense. These update procedures, however, require the tuning of many hyperparameters.

Here, we propose a novel and efficient update technique, optimal in the least-square sense, that follows the two-step training procedure of the off-line training. First, we obtain the new units in the hidden layer by performing a sequential \textit{k-means} clustering of the observations detected outside the training range. Secondly, we efficiently retrain the whole RBF in "one-go" using the Schur complement.

\subsubsection{Online \textit{k-means}}
The fist step in the update procedure is to generate new centers on the new subspace defined by the data points detected outside of the initial training range. This set of points is denoted as $\mathbf{X}^e \subset \mathbf{X}^t$. During the off-line initial training stage, the tesselation of the latent space was generated using the \textit{k-means} algorithm, resulting in the set of centroids (centers for the RBF) $\mathbf{X}^c$. The count of the number of training points associated with each centroid is saved in matrix $\mathbf{C} \in \mathbb{R}^{N_\textrm{R}}$. 

In the online stage, we use an in-house adapted version of the k-means algorithm for sequential data, namely the \textit{sequential k-means} algorithm, which is close to the original formulation of the \textit{k-means} algorithm of \cite{macqueen1965}, see also \cite{duda2006}. The pseudo-code of the algorithm is described in Alg. \ref{alg:online-k-means}.

\begin{algorithm*}
    \SetKwData{Left}{left}
	\SetKwData{This}{this}
	\SetKwData{Up}{up}
	\SetKwInOut{Input}{input}\SetKwInOut{Output}{output}

	\For{$\mathbf{x}^e \in \mathbf{X}^e$}{
		Find the closest centroids $\mathbf{x}^c_j \in \mathbf{X}^c$ to $\mathbf{x}^e$  \\
		Compute $r= \lVert \mathbf{x}^e -  \mathbf{x}^c_j \rVert $ \\
		\eIf{$r<d_e$}{
			$C[j] \gets C[j] + 1$ \\
			$\mathbf{x}^c_j \gets \mathbf{x}^c_j  + (\mathbf{x}^e-\mathbf{x}^c_j)/C[j]$ \\
		}{
			Append $\mathbf{x}^e$ to $\mathbf{X}^c$ \\
			$N_\textrm{R} \gets N_\textrm{R} + 1$ \\
			Append $1$ to $\mathbf{C}$ \\	
		} 
	}
	\caption{Online k-means pseudo-code}\label{alg:online-k-means}
\end{algorithm*}

\begin{figure}[htbp]
    \centering

    \subfloat[\label{subfig:alpha1}]{%
      \includegraphics[width=0.45\textwidth]{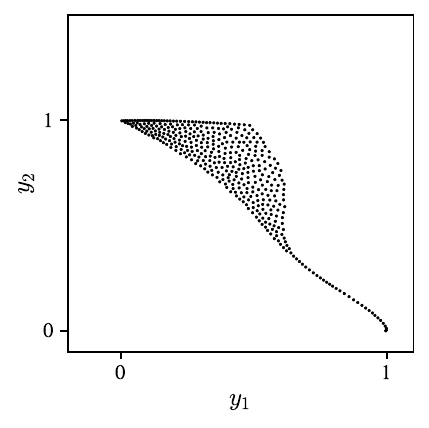}
    }
    \subfloat[\label{subfig:alpha10}]{%
      \includegraphics[width=0.45\textwidth]{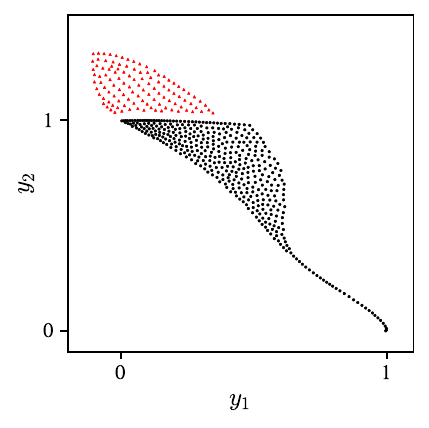}
    }
    \caption[Demonstration of the online \textit{k-means} algorithm.]{(a) Initial tesselation of the low-dimensional space after applying the \textit{k-means} algorithm with $N_\textrm{R} = 200$ where each circular dot represents a cluster centroids $\mathbf{y}^c \in \mathbf{Y}^c$. (b) Application of the online \textit{k-means} algorithm on the set of states detected in extrapolation, $\mathbf{X}^e$ (blue region in Fig. \ref{fig:extrapolation}, generating 55 new centroids, represented by red triangles.}\label{fig:online_kmeans}
\end{figure}

A clear advantage of the distance-based formulation is that the numbers of new centers is automatically determined by the algorithm and does not come as an extra hyper-parameter. The algorithm is applied on the set $\mathbf{X}^e$ (blue points of Figure \ref{fig:extrapolation}), illustrated in Figure \ref{fig:online_kmeans}.

\subsubsection{Update of the RBF weights}
Considering a RBF with gaussian kernel, as long as all centers are distinct, the kernel matrix is always a symmetric positive definite real matrix and is therefore invertible. Let $\mathbf{\Phi}_{1,1} \in \mathbb{R}^{n\times n} $ represent the initial kernel matrix, $\mathbf{\Phi}_{2,2} \in \mathbb{R}^{m\times m}$ the kernel matrix of the new centers added by the online k-means algorithm, and $\mathbf{\Phi}_{1,2} = \mathbf{\Phi}_{2,1}^T \in \mathbb{R}^{n\times m}$ the cross kernel matrix between initial and new centers, respectively. Optimally updating the model parameters in the least-square sense requires the inversion of the augmented kernel matrix, of size $(n+m)\times(n+m)$, defined as, 

\begin{equation}
	\mathbf{\Phi} = \left( \begin{array}{c c}
	\mathbf{\Phi}_{1,1} & \mathbf{\Phi}_{1,2} \\
	\mathbf{\Phi}_{2,1} & \mathbf{\Phi}_{2,2} \\
	\end{array} \right),
	\label{eq:augmented_kernel}
\end{equation} 

As the number of centers increases, this task can become computationally intensive, especially for an on-line procedure with high retrain frequency. To that end, we use the Schur complement for an efficient matrix inversion.

Let $M$ be a square matrix of size $(n+m)\times(n+m)$, written in terms of block partitions as
\begin{equation}
	M = \left( \begin{array}{c c}
	A & B \\
	C & D \\
	\end{array} \right),
\end{equation} 
where $A \in \mathbb{R}^{n\times n}$, $B \in \mathbb{R}^{n\times m}$, $C \in \mathbb{R}^{m\times n}$ and $D \in \mathbb{R}^{m\times m}$. If $A$ is invertible, the Schur complement of block $A$ of matrix $M$ is defined as 
\begin{equation}
	M/A = D-CA^{-1}B.
\end{equation}
We then have the following theorem (see \cite{gallier2011} for a proof):

\begin{theorem}[Invertibility of Schur complement]
\label{pythagorean}
If $A$ is invertible, then \\

\begin{center}
$M$ is invertible $\Longleftrightarrow$ $M/A$ is invertible, \\
\end{center}

which implies that 

\begin{equation}
	M^{-1} = \left( \begin{array}{c c}
	A^{-1} + A^{-1} B (M/A)^{-1} C A^{-1} & -A^{-1}B(M/A)^{-1} \\
	-(M/A)^{-1}CA^{-1} & (M/A)^{-1} \\
	\end{array} \right).
	\label{eq:schur_inverse}
\end{equation}
\end{theorem}

\begin{figure}[htbp]
    \centering
    \includegraphics[width=0.9\textwidth]{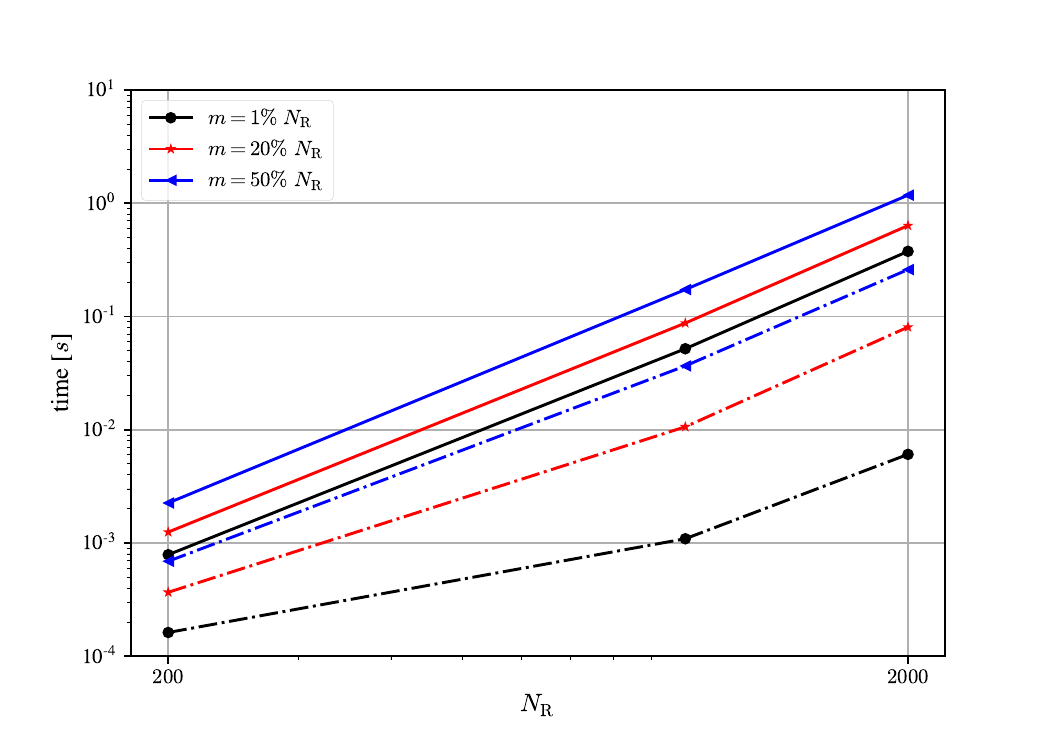}
    \caption[Time complexity of RBF update procedure.]{Comparison of the retraining run-time of a RBF with $N_\textrm{R}$ numbers of centers initially and $m$ (defined as percentage of $N_\textrm{R}$) new centers. Solid lines correspond to direct inversion and dashed-dotted lines to the technique using the Schur complement.}
    \label{fig:time_schur}
\end{figure}

Based on the definition of the augmented kernel matrix, Eq.~(\ref{eq:augmented_kernel}), it is straightforward to apply the Schur complement for efficient matrix inversion. In fact, since $\mathbf{\Phi}_{1,1}^{-1}$ is known from the initial RBF training and $\mathbf{\Phi}$ is invertible (as a symmetric positive definite real matrix), theorem 1 states that the Schur complement of the block $\mathbf{\Phi}_{1,1}$ of matrix $\mathbf{\Phi}$ is invertible. 
We can then compute its inverse and immediately construct $\mathbf{\Phi}^{-1}$ with Eq.~(\ref{eq:schur_inverse}). Hence, inverting the whole matrix requires only the inversion of a $m \times m$ matrix instead of a $(n+m)\times(n+m)$ one. This drastically reduce the retraining time of the RBF, as shown in Figure \ref{fig:time_schur}, where run-time is compared with that of the direct inversion. When $m$ is proportionally small compared to $n$, direct inversion is one to two order of magnitudes slower. During a simulation, since $f_\textrm{retrain}$ is set to a low value, $m$ is always small compared to $n$ (less than 5\% of $m$ in practice). This provides efficient inversion of the RBF system matrix while maintaining optimality of the model in the least-square sense.

\begin{figure}
    \centering
     \includegraphics[width=0.8\columnwidth]{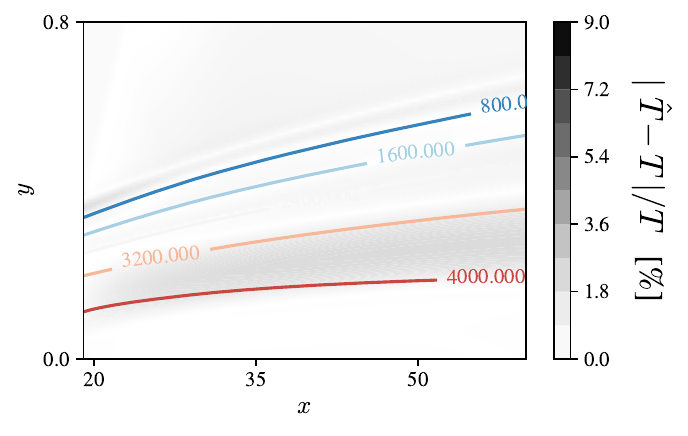}%
    \caption[Demonstration of the full update procedure on case A.]{Relative temperature error $\lvert \hat{T}-T \rvert/T$ in percent with contours of temperature in the physical space after retraining.}\label{fig:retrain_full_scale}
\end{figure}

\begin{figure}
\begin{center}
    \includegraphics[width=0.9\columnwidth]{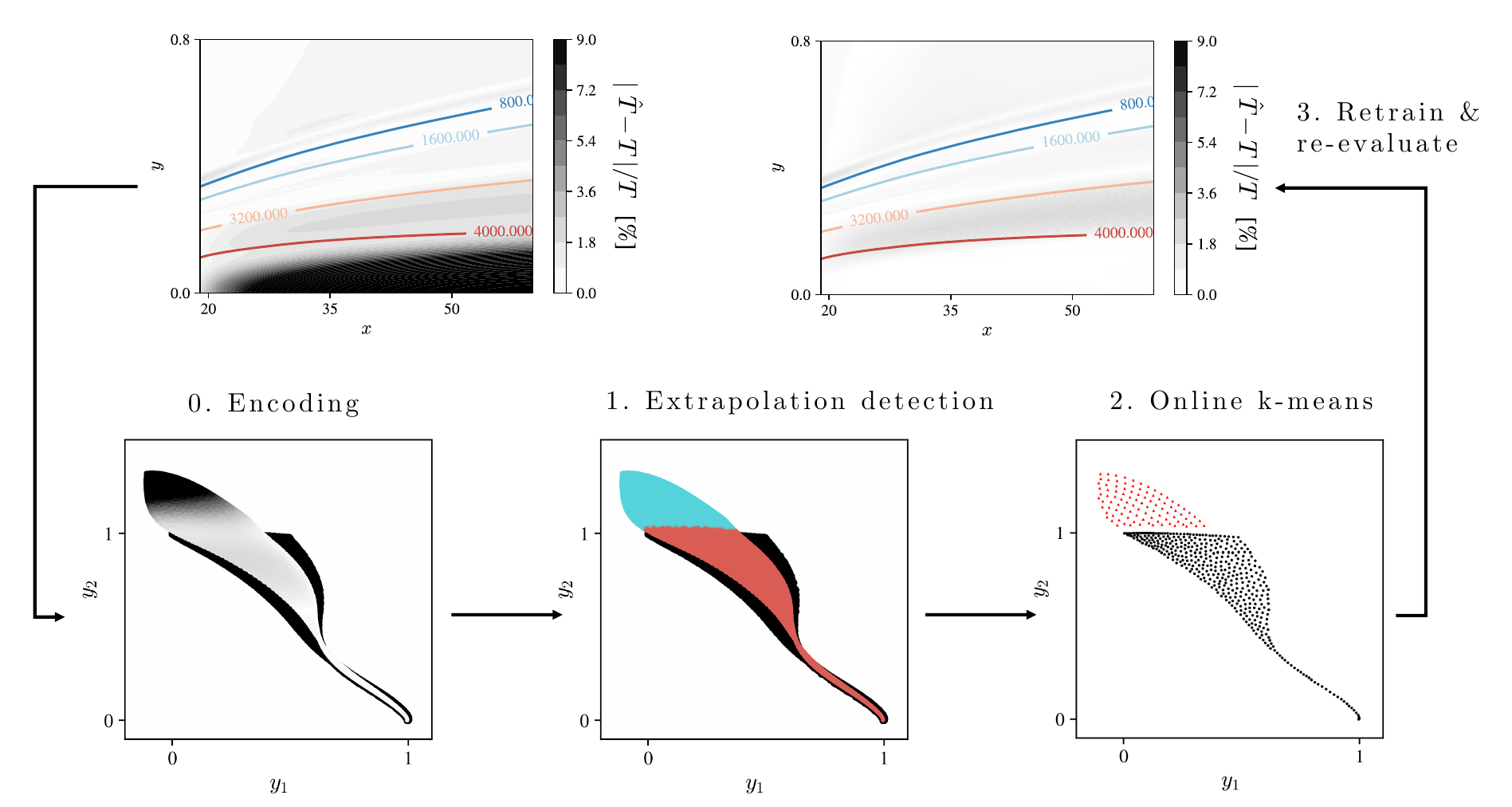}
        \caption[]{Illustration of the complete retraining procedure of the model in an online scenario.}
        \label{sketch_active}
	\end{center}
\end{figure}

Finally, the RBFs corresponding to each cluster are retrained and the resulting error in physical space with the true value of Mutation++ are shown in Figure \ref{fig:retrain_full_scale}. Error drastically decreases in the extrapolation region, demonstrating the efficient and optimal active learning capability of the model. The complete procedure is sketched in Figure \ref{sketch_active}.

\section{Results}
\label{tests}

This section presents the application of the RONAALP algorithm to three different numerical simulations of hypersonic flows in chemical nonequilibrium.

\subsection{Flow configuration}
The flow configuration chosen to showcase the technique is the adiabatic flat-plate boundary layer in Earth's atmosphere at $Ma = 10$, based on \cite{Marxen2013,Marxen2014a}. The freestream and thermodynamic conditions are presented in Table \ref{table:condition_olaf}. The origin of the coordinate system is placed at the (virtual) leading edge of the flat plate. The domain simulated extends in the streamwise direction from $x=14$ to $x=85$ non-dimensional units with 960 points, equally spaced. At the inflow, a self-similar solution in chemical non-equilibrium is prescribed, \citep{lees1956, williamsCTR2021}. Starting at $x=70$, the solution is damped to the self-similar reference solution using a numerical sponge. The wall-normal direction is discretized from $y=0$ to $y=1.6$ using 211 grid points clustered near the wall with a cuboid stretching function. The last 26 points in the freestream are also assigned to a sponge layer. Finally, periodicity is assumed in the spanwise direction.

\begin{table}[htbp]
\begin{center}
\begin{tabular}{c l r}
 \multicolumn{3}{c}{Test case} \\
 \hline
  \hline
 $ M_\infty $ & & 10 \\ 
 $ Re_\infty $ & & $10^5$ \\ 
 $ T_\infty$    &     [K]  & 350  \\ 
 $\rho_\infty$  &   [kg/m$^3$] & $3.56 \times 10^{-2}$ \\ 
 $ p_\infty$    &    [Pa]  & 3596 \\ 
 $ c_\infty$    &    [m/s] & 375.41 \\
 $ L_{ref}$   &  [m] & $1.6 \times 10^{-2}$  \\
 \hline
 \hline
\end{tabular}
\caption{Thermodynamic, freestream and forcing conditions for the Mach-10 adiabatic flat-plate boundary layer with blowing and suction test case.}
\label{table:condition_olaf}
\end{center}
\end{table}

\subsection{Transient simulation}
The first test chosen is a low-fidelity to high-fidelity transient simulation. In fact, after the initialization of the flow with a self-similar solution (low-fidelity solution), the simulation transiently evolves until converging to the steady solution of the Navier-Stokes equations (high-fidelity solution or baseflow). Large differences are observed between the low-fidelity and high-fidelity solutions using the full-order thermochemical gas model. Hence, a model trained only on the low-fidelity simulation will encounter many new states that have to be actively learned throughout the transient simulation, as shown in the previous section. 

In the initial study of \cite{scherding2023data}, the reduced-order thermochemical model was proven to maintain a stable high-fidelity baseflow while speeding up the evolution of thermochemical properties by up to 70\%. As no features were added and the model remained stable, the evaluation of the thermochemical properties remained well within the training interval (sampled from the baseflow) of the model where high accuracy is guaranteed. However, a test using a model trained on the low-fidelity solution without the active learning procedure showcased numerical instabilities and crashed (not shown here). This highlights again the need for an online update procedure.

The base model is trained off-line on the locally self-similar solution, and white noise with 2\% amplitude is added to improve the robustness of the model. The hyper-parameters of the models are : $d=2,~ N_C=2, ~N_R = 200$. During the simulation, the model is updated every 100 iterations.

\begin{figure}
    \centering

    \subfloat[]{%
      \includegraphics[width=0.25\textwidth]{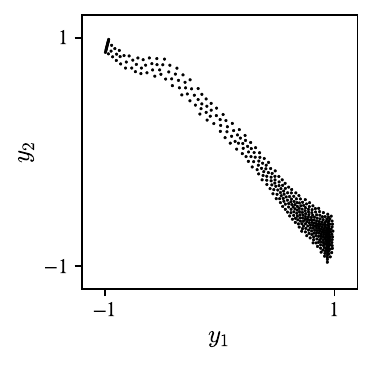}
    }
    \subfloat[]{%
      \includegraphics[width=0.75\textwidth]{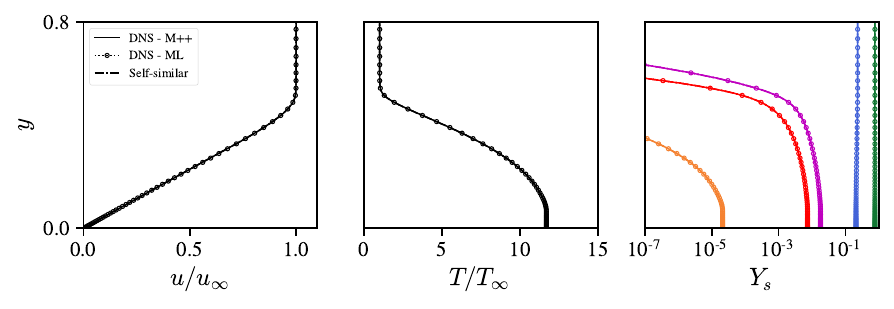}
    }
    \\
   	    \subfloat[]{%
      \includegraphics[width=0.25\textwidth]{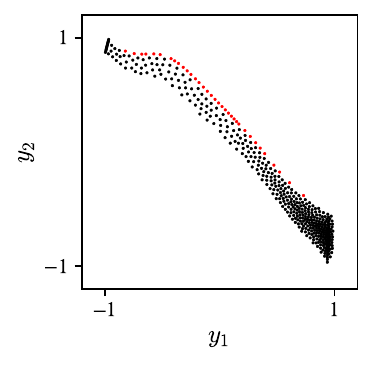}
    }
    \subfloat[]{%
      \includegraphics[width=0.75\textwidth]{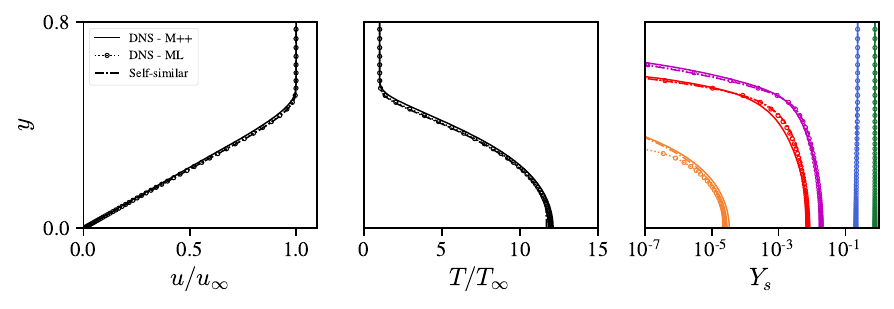}
    }
    \\
   	    \subfloat[]{%
      \includegraphics[width=0.25\textwidth]{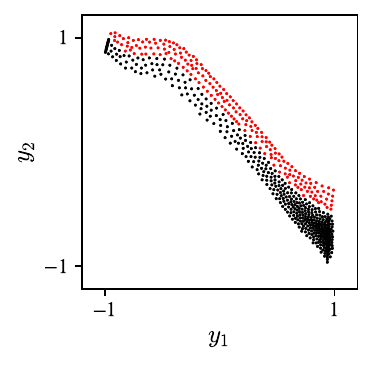}
    }
    \subfloat[]{%
      \includegraphics[width=0.75\textwidth]{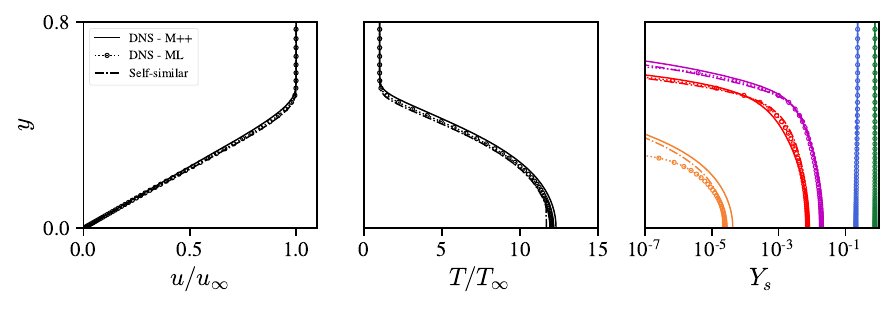}
    }
    \\
   	    \subfloat[]{%
      \includegraphics[width=0.25\textwidth]{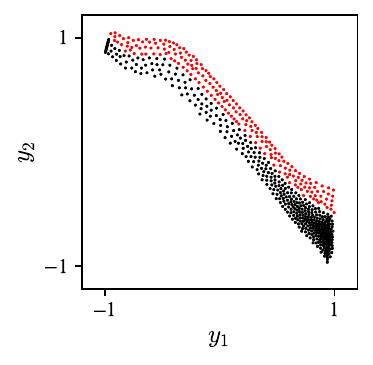}
    }
    \subfloat[]{%
      \includegraphics[width=0.75\textwidth]{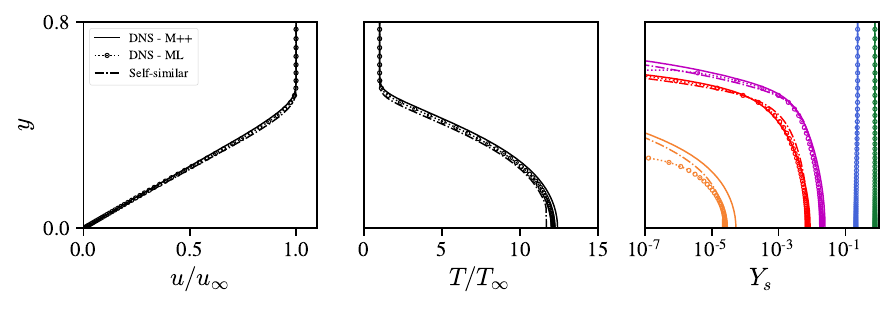}
    }
    \\
   	    \subfloat[]{%
      \includegraphics[width=0.25\textwidth]{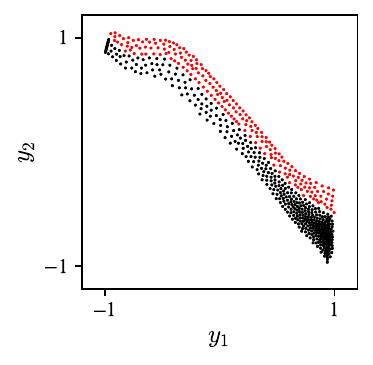}
    }
    \subfloat[]{%
      \includegraphics[width=0.75\textwidth]{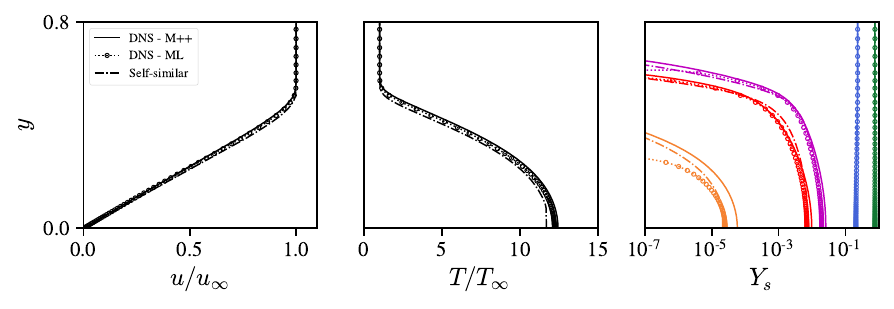}
    } 
        \phantomcaption 
\end{figure}
      
\begin{figure}
	\ContinuedFloat
    \centering
    
    \subfloat[]{%
      \includegraphics[width=0.25\textwidth]{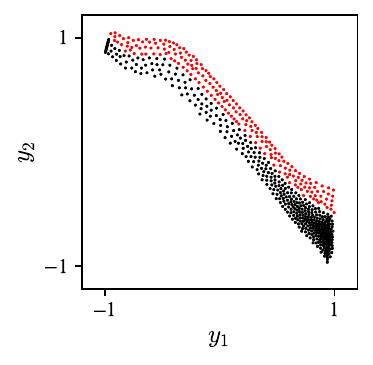}
    }
    \subfloat[]{%
      \includegraphics[width=0.75\textwidth]{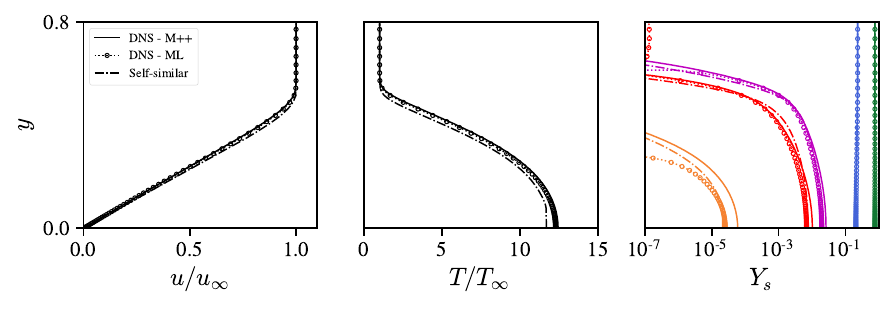}
    }  

   	    \subfloat[]{%
      \includegraphics[width=0.25\textwidth]{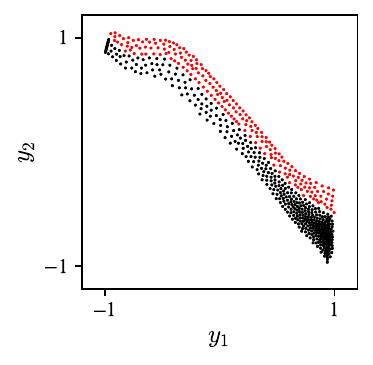}
    }
    \subfloat[]{%
      \includegraphics[width=0.75\textwidth]{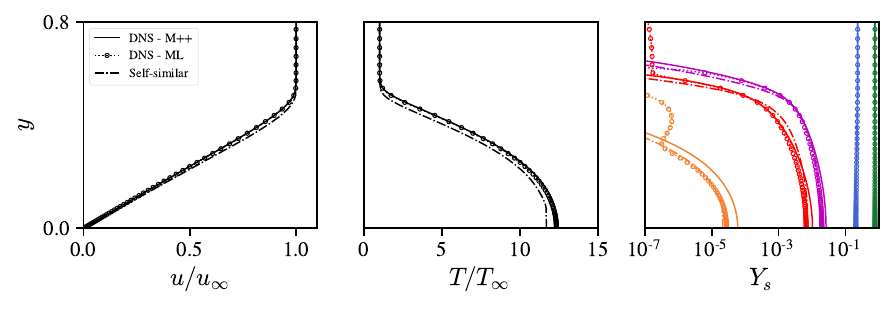}
    }
    \\
   	    \subfloat[]{%
      \includegraphics[width=0.25\textwidth]{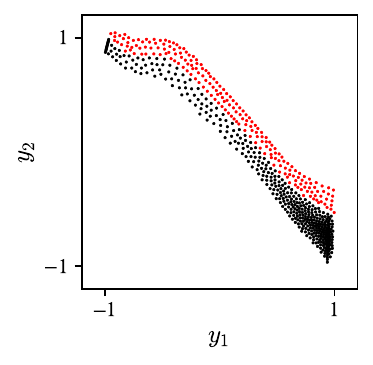}
    }
    \subfloat[]{%
      \includegraphics[width=0.75\textwidth]{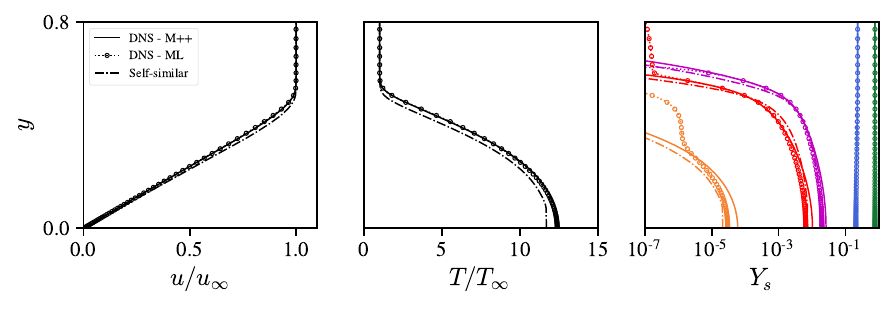}
    }
    \\
   	    \subfloat[]{%
      \includegraphics[width=0.25\textwidth]{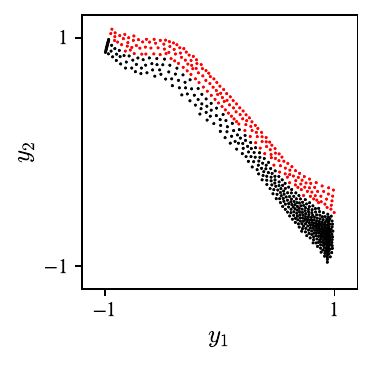}
    }
    \subfloat[]{%
      \includegraphics[width=0.75\textwidth]{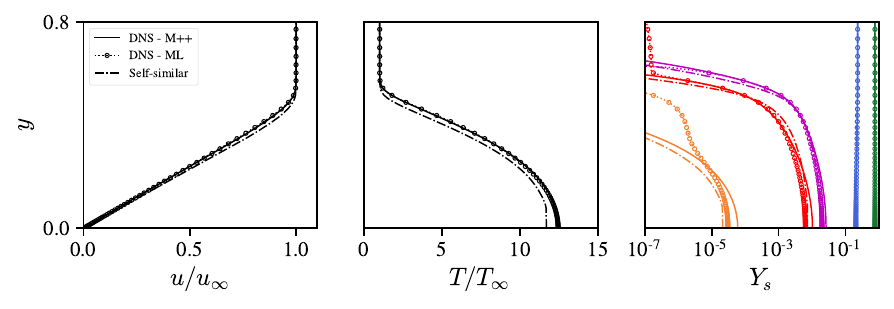}
    }    
    \phantomcaption 
\end{figure}
      
\begin{figure}
	\ContinuedFloat
   	    \subfloat[]{%
      \includegraphics[width=0.25\textwidth]{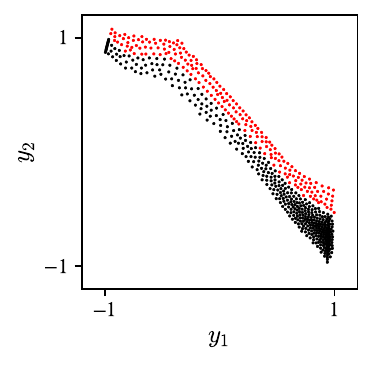}
    }
    \subfloat[]{%
      \includegraphics[width=0.75\textwidth]{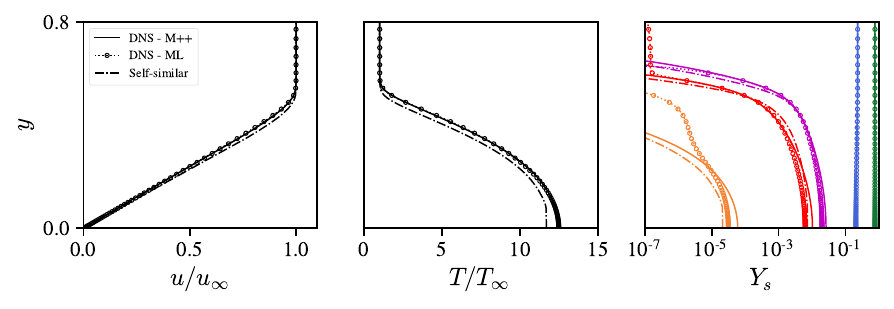}
    }    
    
    \caption[Transient simulation of case A using full and reduced-order thermochemical models.]{Transient low to high-fidelity simulation of case A using both full and reduced-order thermochemical models. Left column: RBF centers in the latent space. Black and red dots represent initial and newly added centers during the simulation, respectively. Right column: evolution of the boundary layer profiles. Solid and dotted lines with markers correspond to the solution using the full and reduced-order thermochemical model, respectively. Each line represents uniformly sampled instantaneous snapshot at a non-dimensional time $t \in [0,10]$, ordered chronologically.}\label{fig:active_caseA}
\end{figure}

Figure \ref{fig:active_caseA} presents the evolution of the population of the RBF centers in the latent space as well as boundary layer profiles at $x=40$ at different instant during the simulation. As the simulation advances in-time, more centers are added as new thermodynamic states are encountered. It is important to note that the number of centers rapidly plateaus after a first transient where many new states are discovered. The final total number of centers is 602. This proves the convergence of the active learning process during the simulation. 
Correspondingly, the boundary layer profiles generated with RONAALP converge closely to the reference simulation using Mutation++. The main difference is in the radical mass fractions. However, they are present in such small quantities that they do not alter more relevant quantities of interest such as the velocity profiles and maximum temperature within the boundary layer. Furthermore, a reduced-order model is not expected to be precise in the range $[0,10^{-5}]$.

%

\begin{figure}[htbp]
    \centering
    \includegraphics[width=1.\textwidth]{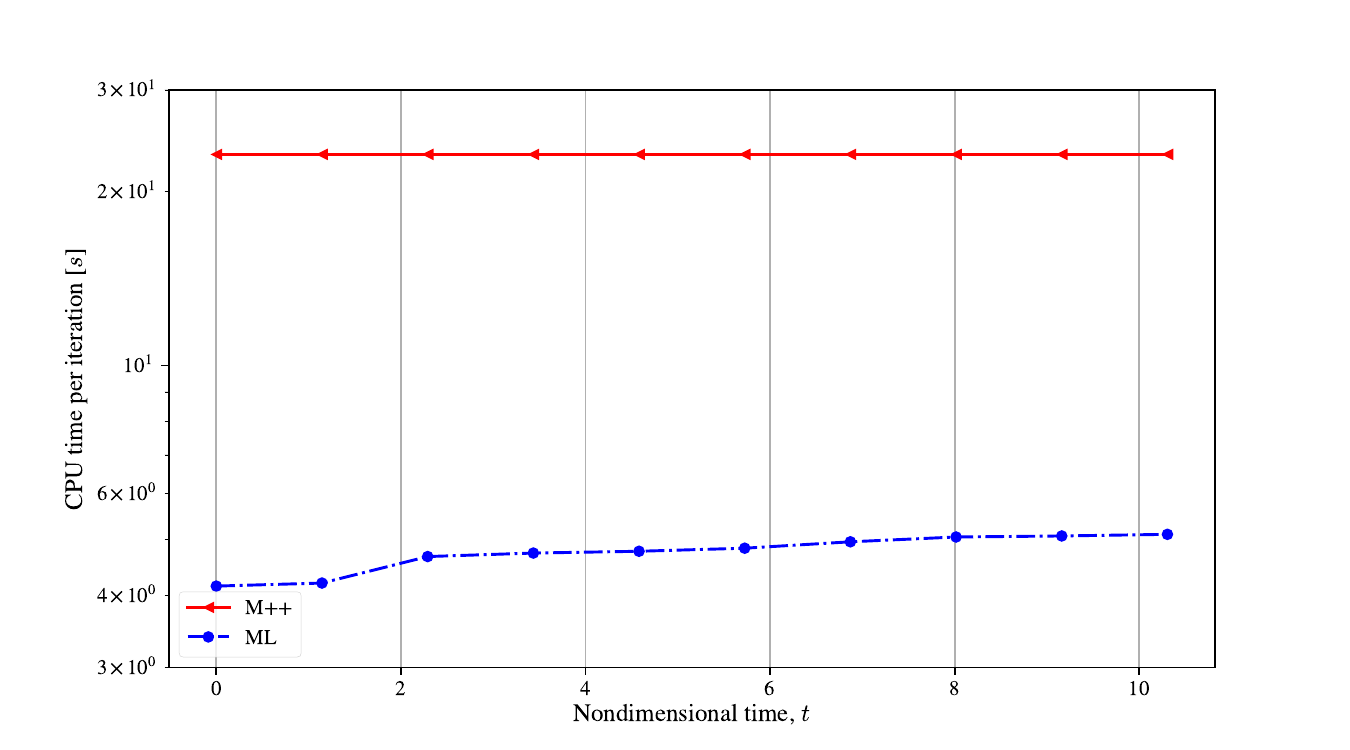}
    \caption[Time per iteration during transient simulation of case A.]{Comparison of time per iteration during the transient simulation using both full (solid line) and reduced-order thermochemical model (dashed-dotted).}\label{fig:time_iter_A}
\end{figure}

Figure \ref{fig:time_iter_A} presents the evaluation time of thermochemical properties for all grid points. The model initially performs 80\% faster than Mutation++. As the solution progresses and new centers are added, the performance of the model slightly degrades until reaching a final performance that is 75\% faster than Mutation++. This loss of performance is due to the higher evaluation cost after progressively growing the RBF. In fact, we recall here that the time complexity of the evaluation step of the RBF is $O( C_{RBF} \times N_t \times N_\textrm{R} \times d)$. Hence, as $N_\textrm{R}$ increases during the active learning process, so does the time complexity. However, since the number of new centers at each update is small and the load is split between two clusters, this additional cost is not detrimental to the overall performance of the data-driven model. 

The time spent during the updating step is however hard to evaluate since it depends on the number of points detected during extrapolation. To evaluate it empirically, two simulations have been run for 1000 iterations, with and without update every 100 iterations, respectively. The simulation with updates was 1.05 times slower, even though it was initialized with the locally self-similar solutions and the update load is higher early on during the transient, as seen on Figures \ref{fig:active_caseA} and \ref{fig:time_iter_A}. Hence, the update time is marginal as compared to the total time to solution and can be omitted as a first approximation. Hence, directly integrating the curve of the data-driven reduced order model leads to a time to solution 77\% times faster than Mutation++.

This first test proves that RONAALP can actively learn new input/output relations on-the-fly. Thus, the models are predictive, generalizable and can help reduce the high cost associated with high-fidelity simulation of hypersonic flows in chemical nonequilibrium, with minimal impact on the accuracy of the solution. 

\subsection{Optimally disturbed boundary layer with blowing-suction}
The second test chosen is the addition of new dynamics in the baseflow. 
In fact, new unsteady flow features can lead to thermodynamic and composition states outside of the training range. In both cases, extrapolation of the model is required which, if not handled properly, can lead to error build-up over time and will ultimately alter the dynamics in the boundary layer.

To showcase that RONAALP can deal with such events, a single frequency, two-dimensional disturbance is introduced in the baseflow at the wall with blowing and suction on a strip extending from $x_s=19.3$ to $x_e=20.7$, centered at $x_c=20$. This disturbance can be classified as a second-mode instability according to \cite{Malik1991}, which is most amplified in high speed boundary layer flows \citep{mack1975,mack1984}. This case was initially studied in \cite{Malik1991} using linear stability theory and the non-dimensional frequency of $\omega=3.4 \times 10^{-5}$ was predicted to be the most amplified. This case was later revisited \cite{Marxen2011,Marxen2013} using direct numerical simulations using different thermochemical models (perfect gas, chemical equilibrium and nonequilibrium). Good agreement for the growth-rate and amplitude functions were found at $R=\sqrt{Re_x} = 2000$ compared to the earlier results. The analysis was then further extended to weakly nonlinear stages in \cite{Marxen2014a}. The same direct numerical simulation case (without the weakly nonlinear analysis) was reproduced with the present solver and validated against the results of Marxen \textit{et al.} in chemical non-equilibrium using Mutation++ \citep{margaritis2022}.  

The disturbance has a nondimensional forcing frequency $\omega$ given in Eq. (\ref{eq:disturbance}).
\begin{equation}
    \omega = 2\pi\tilde{f} \frac{\mu_\infty}{\rho_\infty {u_\infty}^2}.
    \label{eq:disturbance}
\end{equation}

Here $\tilde{f}$ is the dimensional frequency of the disturbance. The amplitude of the velocity perturbation $A$ is defined as a fraction of the freestream velocity $u_\infty$. The freestream conditions, reference scales, forcing frequency and amplitude are summarized in Table \ref{table:condition_olaf}. The velocity boundary conditions are defined as in \cite{Marxen2011,Marxen2013} and read
\begin{equation}
    \begin{array}{l}
        v(x,0,t) = A sin(\omega t) s(\xi) \\
        u(x,0,t) = 0.
    \end{array}
\end{equation}
The shape function $s$ is defined within the strip as 
\begin{equation}
    s(\xi) = 18.1875\xi^5 - 35.4375\xi^4+20.25\xi^3,
\end{equation}
where the auxiliary coordinate $\xi$ is defined as
\begin{equation}
    \xi = \left\{
    \begin{array}{lcl}
        (x-x_s)/(x_c-x_s) &\phantom{1234}& \mbox{for} ~ x_s<x<x_c,  \\
        (x-x_e)/(x_e-x_c)  &\phantom{1234}& \mbox{for} ~ x_c<x<x_e,  \\
        0 && \mbox{otherwise.}
    \end{array}
\right.
\end{equation} 
The unsteady simulation is then advanced until transient effects are advected out of the domain and a time-periodic state is achieved.

In order to compare the dynamics of the unsteady simulations, $N_T=100$ flow snapshots are collected over one forcing period and Fourier transformed in time $t$. This yields a Fourier coefficient $\hat{\phi}_{hrm}$ for a given primitive quantity $\phi \in [\rho, u, v, w, P, T, \ldots]$ and harmonic $hrm$. Since the disturbance introduced has a single frequency, only the results for the first harmonic ($hrm=1$) will be compared and the subscript will be omitted. 

In the following, the streamwise disturbance amplification obtained with either Mutation++ or RONAALP will be compared using: i) the pressure disturbance at the wall, $\hat{p}_{wall}$ (equivalent to the RMS wall pressure), and, ii) the wall-normal maxima of the streamwise velocity disturbance $\hat{u}_{\textrm{max}}$ defined as 
\begin{equation}
    \label{eq:umax}
    \hat{u}_{\textrm{max}}(x) = \max_{y} (\hat{u}(x,y)),
\end{equation}
as it is common practice in the literature \citep{Marxen2013,Marxen2014a}.

The comparison of the resulting disturbance streamwise amplification consists of a robust test of the accuracy of RONAALP. Indeed, we will be comparing first order statistics that are highly sensitive to any change in the flow properties such as boundary layer height and local Reynolds number. 

\subsubsection{Results}
The model is trained on the steady baseflow solution, supplemented with white noise of 2\% amplitude to increase robustness. The hyper-parameters of the model are : $d=2,~ N_C=2, ~N_R = 200$. Starting from the steady solution, the simulation is advanced with the data-driven model and an update frequency of 100 iterations is applied until a time-periodic state is reached.  Figure \ref{fig:active_caseA_BS} shows the evolution of the population of the RBF centers in the latent space during the transient phase all the way up to the time-periodic state, showing how the model learns online during the simulation.

\begin{figure}
    \centering

    \subfloat[]{%
      \includegraphics[width=0.45\textwidth]{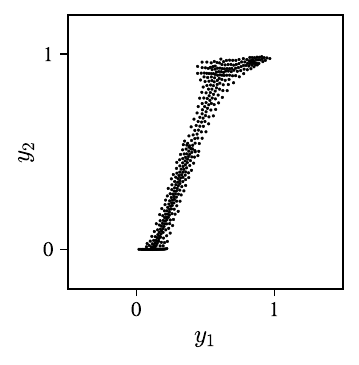}
    }
    \subfloat[]{%
      \includegraphics[width=0.45\textwidth]{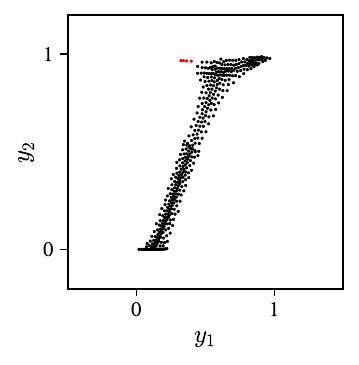}
    }
    \\
    \subfloat[]{%
      \includegraphics[width=0.45\textwidth]{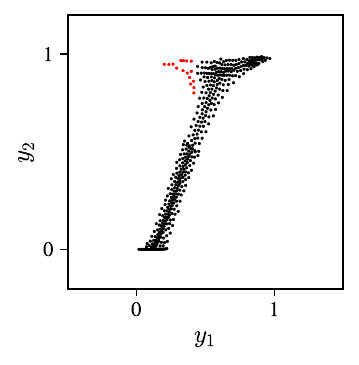}
    }
    \subfloat[]{%
      \includegraphics[width=0.45\textwidth]{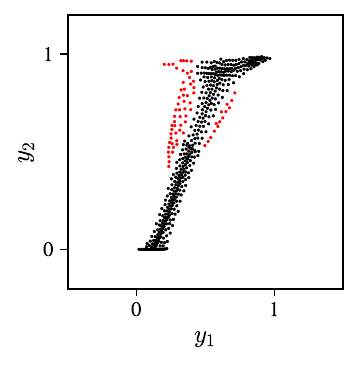}
    }
    \\
    \subfloat[]{%
      \includegraphics[width=0.45\textwidth]{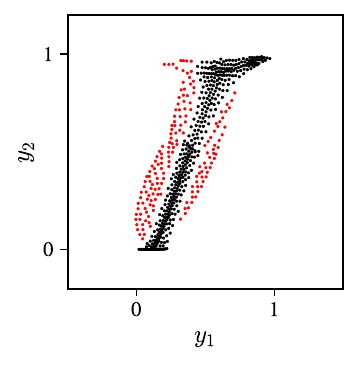}
    }
    \subfloat[]{%
      \includegraphics[width=0.45\textwidth]{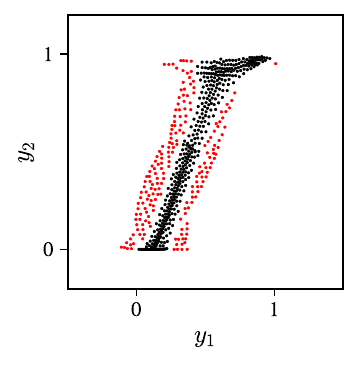}
    }
     \phantomcaption 
\end{figure}
      
\begin{figure}
	\ContinuedFloat
    \centering
    \subfloat[]{%
      \includegraphics[width=0.45\textwidth]{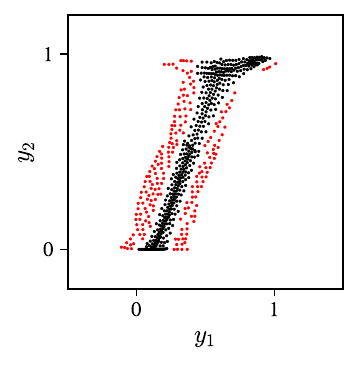}
    }
    \subfloat[]{%
      \includegraphics[width=0.45\textwidth]{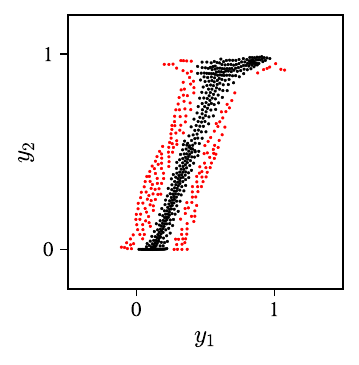}
    }
     \\
    \subfloat[]{%
      \includegraphics[width=0.45\textwidth]{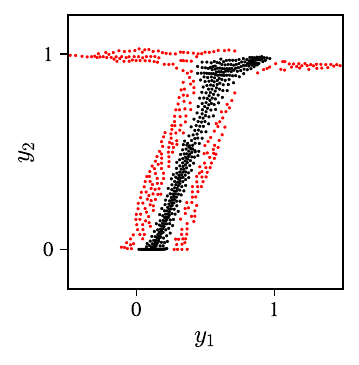}
    }
    \subfloat[]{%
      \includegraphics[width=0.45\textwidth]{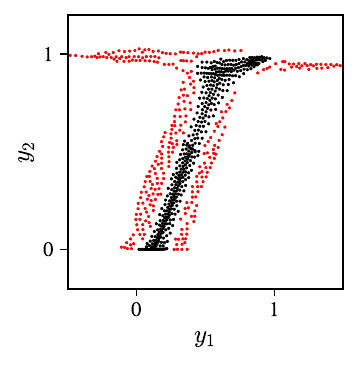}
    }
    \caption[Evolution of RBF centers during a second-mode instability growth simulation.]{Evolution of RBF centers in the latent space during the unsteady simulation of second-mode growth in case A boundary layer. Black and red dots represent initial and newly added centers during the simulation, respectively.}\label{fig:active_caseA_BS}
\end{figure}

\begin{figure}[htbp]
    \centering
    \includegraphics[width=1.\columnwidth]{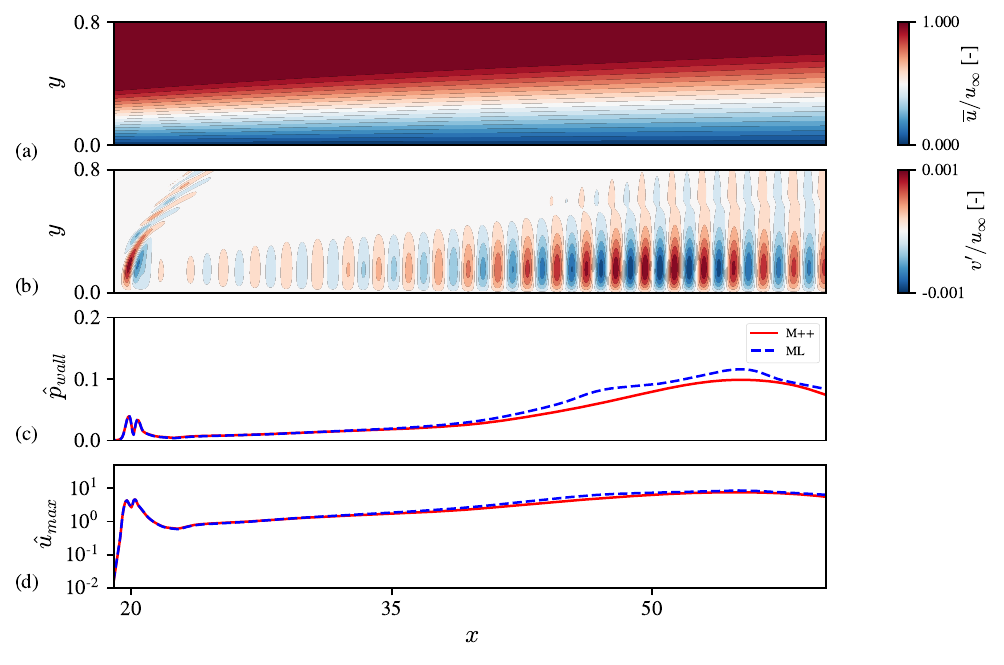}
    \caption[Second-mode instability growth in case A using an adaptive reduced-order thermochemical model.]{Contours of normalized (a) mean streamwise velocity $\overline{u}$, and (b) wall-normal perturbation velocity $v'$. Evolution of (c) wall pressure, and (d) streamwise velocity disturbances, as a function of streamwise position $x$ for case A. Red solid lines correspond to the baseline solution (Mutation++) while blue dashed lines correspond to the result of RONAALP.}\label{fig:growthrate_ML3}
\end{figure}

\begin{figure}[htbp]
    \centering
    \includegraphics[width=1.\textwidth]{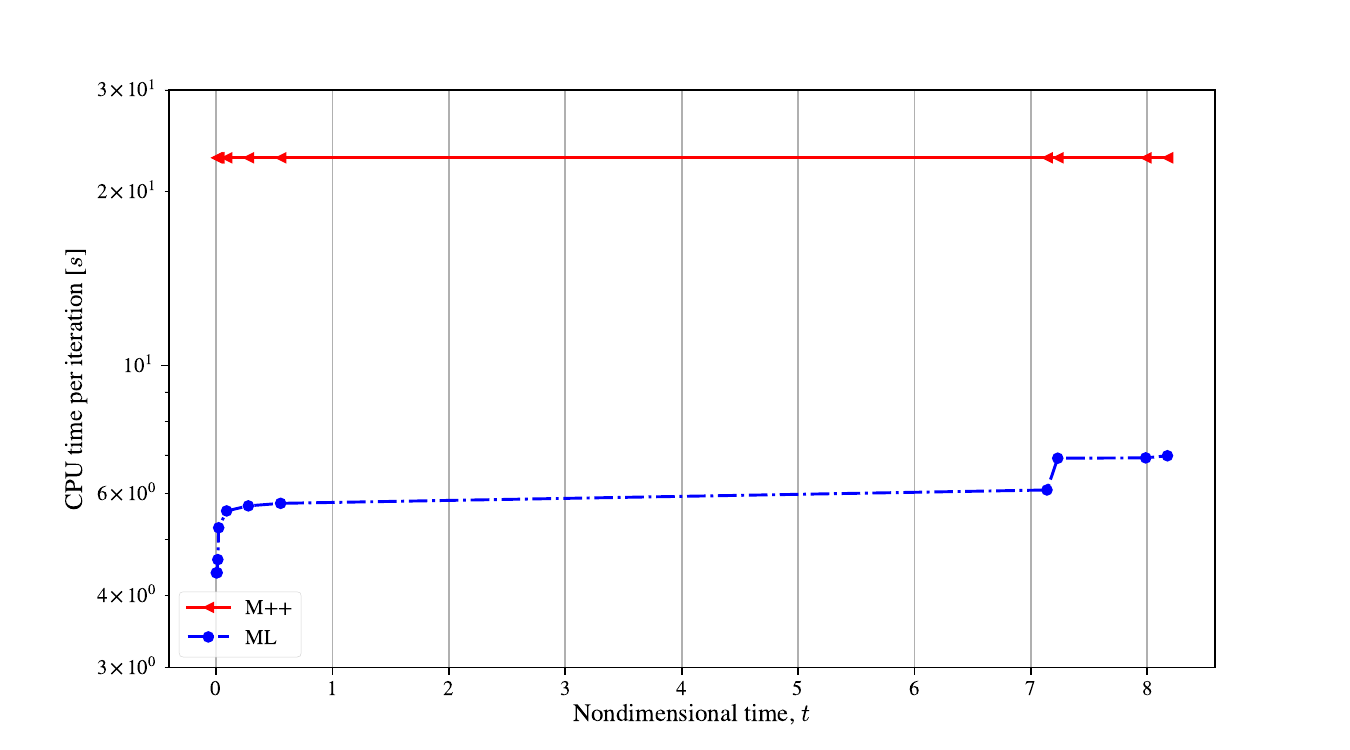}
    \caption[Time per iteration during unsteady simulation of case A.]{Comparison of time per iteration during the unsteady simulation using both full (solid line) and reduced-order thermochemical model (dashed-dotted).}\label{fig:time_iter_A_BS}
\end{figure}

Figure \ref{fig:growthrate_ML3}(c,d) depicts the wall pressure and streamwise velocity disturbances obtained, respectively. The overall dynamic remains fairly close to the reference, with a maximum relative error of 10\% on $\hat{p}_{wall}$, noting that the thermochemical model (i.e TPG or CNEQ) has a much bigger impact on the dynamics, as shown in \cite{margaritis2022}. This demonstrates the algorithm's capability to actively learn a reduced-order thermochemical model that is both accurate and efficient during a time-marching simulation, resulting in decreased CPU time required to obtain reliable results. In fact, Figure \ref{fig:time_iter_A_BS} shows that by integrating the time per iteration, the solution was obtained using 75\% less computational resources than when using Mutation++.
 
\subsubsection{Oblique perturbation in 3D boundary layer}
The third and last test case is a three-dimensional flow based on \cite{Marxen2014a}. The same computational setup is used, extruded in the spanwise direction with a width $L_z=7.85$ using 60 grid points. The perturbation boundary condition at the wall is modified to superpose the same primary two-dimensional waves with oblique waves as follows,
\begin{equation}
    \frac{v}{u_\infty} = \sum_{h=1, k \in (0,2,4)} A_v^{(h,k)} \sin(2\pi\xi -h\omega t +\Phi_0^{(h,k)})\exp(-0.4\xi^2)\cos(k\eta z),
\end{equation}
where $\xi=(x-x_c)/L_{strip}$, $L_{strip}=1.7$, $\eta=2\pi/\lambda_z$ and $\lambda_z = L_z$. The phase shift is set to 0 for the primary wave $\Phi_0^{(1,0)}=0$ and $\Phi_0^{(1,2)}= \Phi_0^{(1,4)} = \pi/4$ for the pair of oblique modes. The amplitudes of the primary mode is set to $A_v^{(1,0)} = 10^{-2}$ while the oblique modes amplitude is two order of magnitude smaller, $A_v^{(1,2)}= A_v^{(1,4)} = 10^{-4}$.

\begin{figure}[htbp]
    \centering
    \includegraphics[width=1.\textwidth]{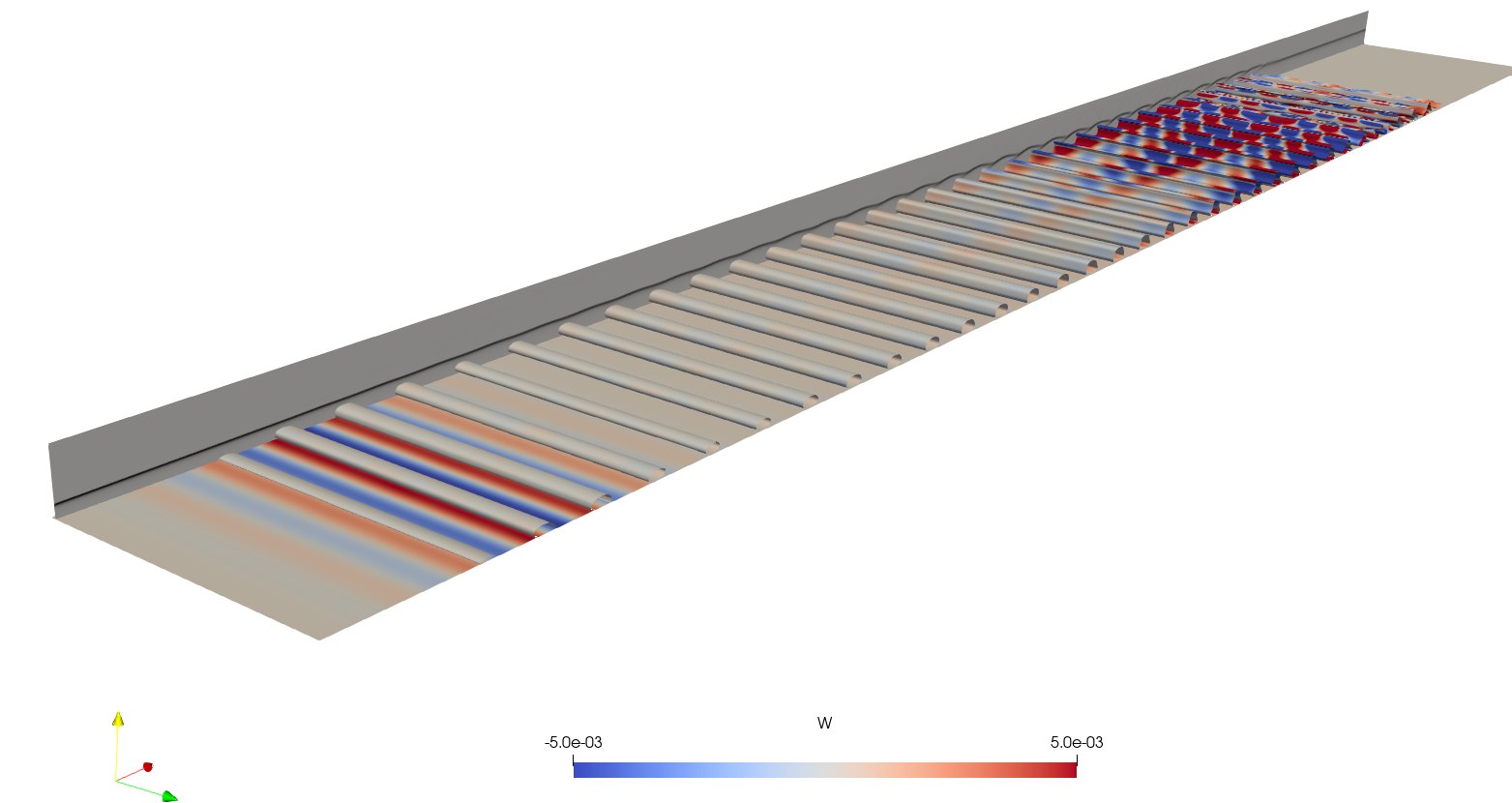}
    \caption[]{Iso-surface of Q-criterion colored by spanwise velocity $w$. Left $x-y$ plane is colored by the normalized magnitude of the density gradient $\lVert \nabla \rho \rVert_2$.}\label{fig:Q_crit_3D}
\end{figure}

The superposition of primary 2D and oblique modes is a commonly used route for simulation of transitional boundary layers \citep{Marxen2014a,passiatore2022, direnzo2021}. Figure \ref{fig:Q_crit_3D} shows the isosurface of the Q-criterion \citep{jeong1995identification}, colored by the spanwise velocity $w$. A numerical schlieren showing the normalized magnitude of the density gradient $\lVert \nabla \rho \rVert_2$ on a two dimensional $x-y$ plane at the left side of the domain is added. The figure illustrates wave interaction, and the emergence of streak structure within the boundary layer.

\begin{figure}
    \centering

    \subfloat[]{%
      \includegraphics[width=0.45\textwidth]{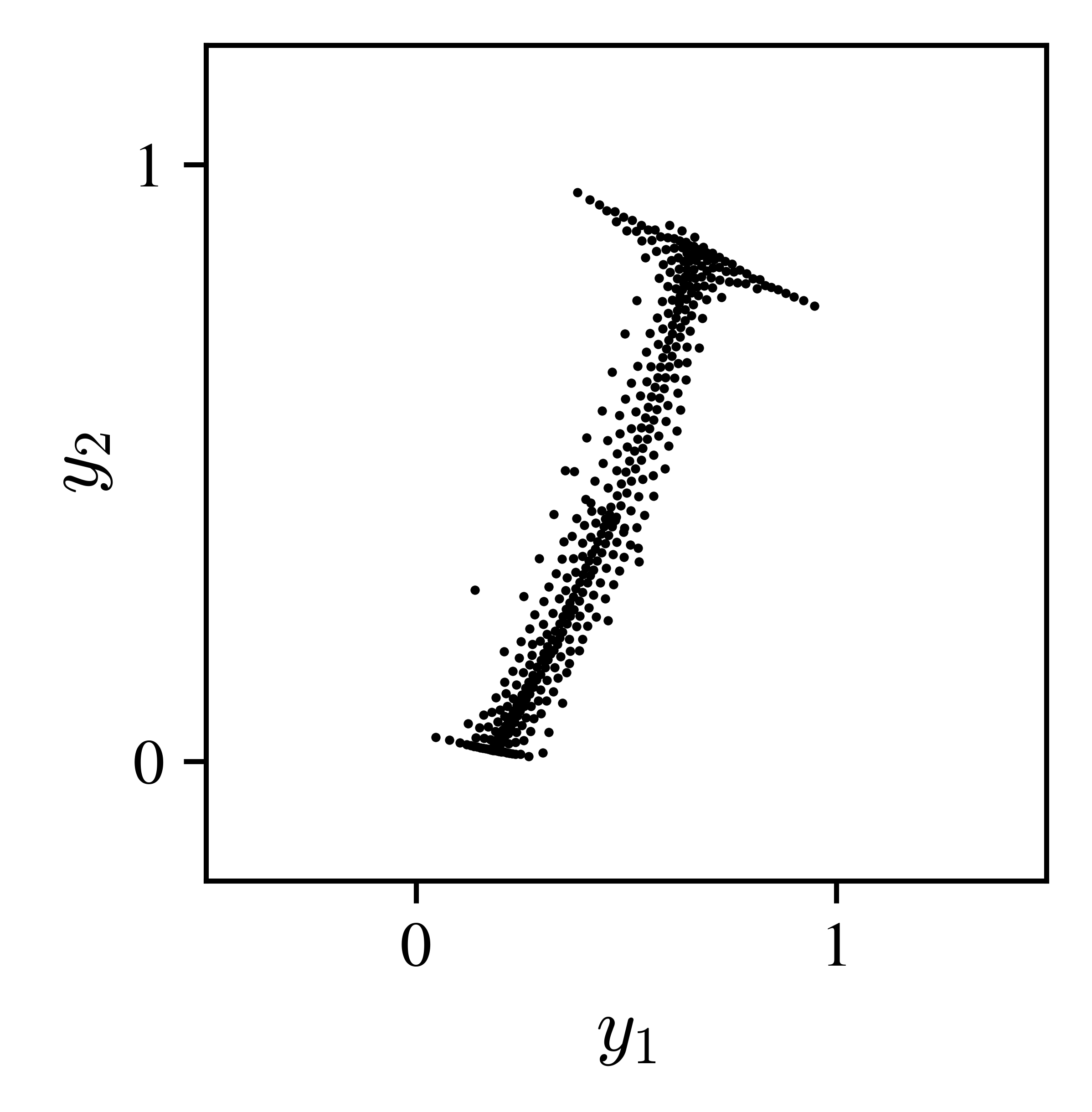}
    }
    \subfloat[]{%
      \includegraphics[width=0.45\textwidth]{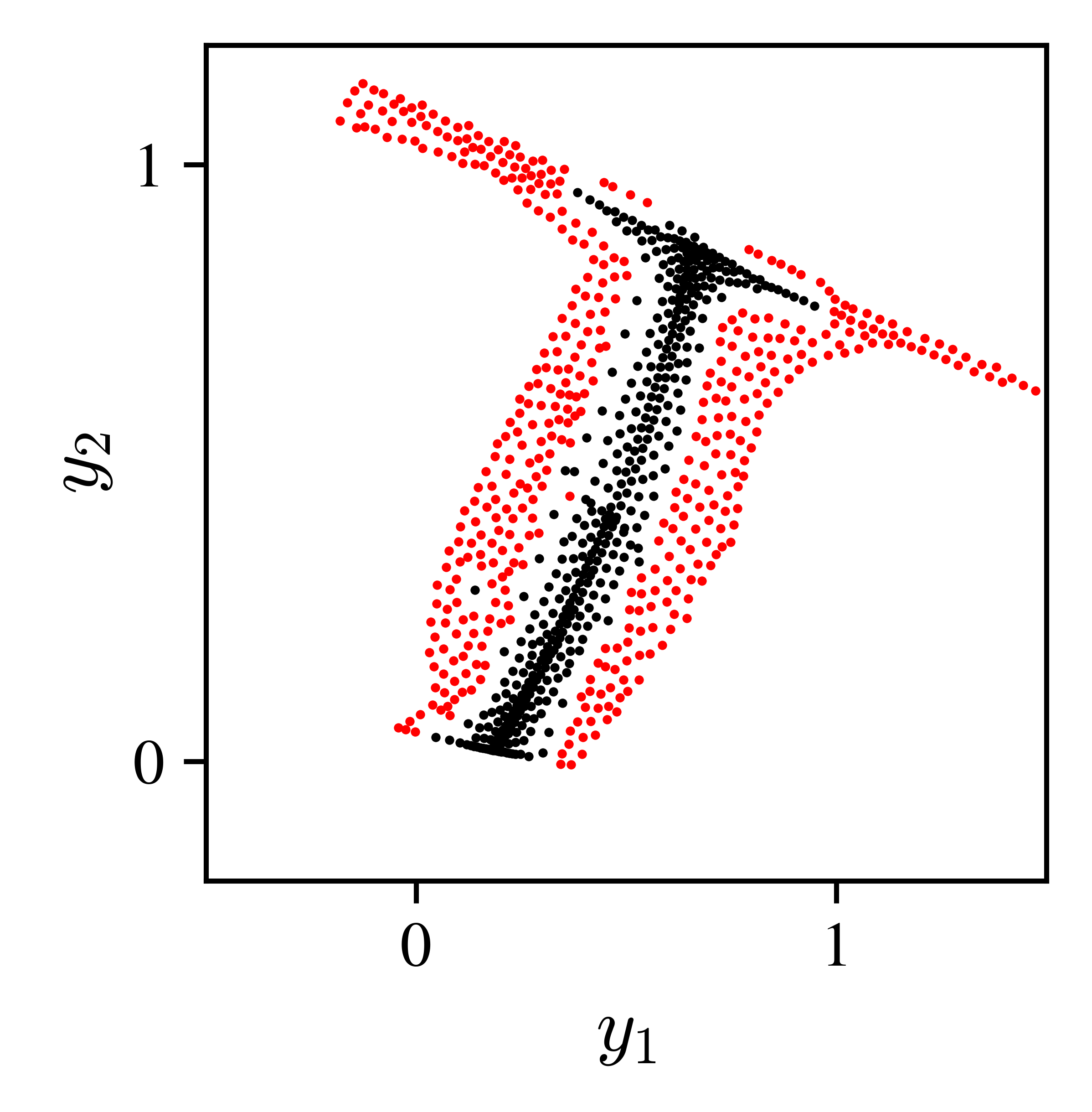}
    }
    \caption{(a) Initial, and (b) final population of RBF centers in the latent space during the simulation of 3D waves interaction. Black and red dots represent initial and newly added centers during the simulation, respectively.}\label{fig:active_3D}
\end{figure}

The model is trained in a curriculum learning fashion using data sampled from a perturbed flow snapshot of the previous two-dimensional unsteady case. The hyper-parameters of the model are again set at $d=2,~ N_C=2, ~N_R = 200$. Starting from the 3D limit-cycle solution obtained with Mutation++, the simulation is advanced with RONAALP (with an update frequency of 500 iterations) until learning convergence (i.e. reaching a steady number of RBF centers). Figure \ref{fig:active_3D}(a,b) shows the evolution of the population of the RBF centers in the latent space during that phase. This highlights that even though the model was trained on a dataset containing thermodynamic states of the optimally disturbed 2D boundary layer, new thermodynamic states pertaining to the three-dimensional waves interaction are generated and actively learned by RONAALP on-the-fly.

\begin{figure}[htbp]
    \centering
    \includegraphics[width=1.\textwidth]{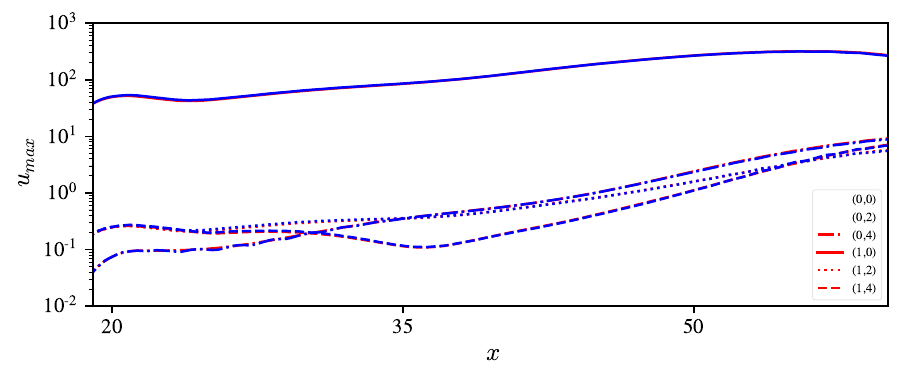}
    \caption[]{Evolution of streamwise velocity disturbances, as a function of streamwise position $x$ for case A. Red lines correspond to the baseline solution (Mutation++) while blue dashed lines correspond to the result of RONAALP. From top to bottom at the left, mode $(1,0)$, $(1,2)$, $(1,4)$ and $(0,4)$.}\label{fig:umax_3D}
\end{figure}

For post-processing, resulting flow snapshots are Fourier transformed in both time $t$ and span $z$. The corresponding Fourier modes are denoted below as $(h,k)$ for a frequency $h\omega$ and spanwise wavenumber $k\eta$. The streamwise velocity disturbance, using the wall-normal maxima (Eq. \ref{eq:umax}), are compared in Figure \ref{fig:umax_3D}. In this three-dimensional case, the dynamics of the simulation that used RONAALP are in almost perfect agreement to the reference solution. Most notably, we observe the emergence of the streak structures (mode $(0,4)$). This demonstrates that RONAALP successfully adapted to account for the three dimensional nature of the flow. 

\section{Conclusion}
\label{conclusion}

In this paper, we presented the RONAALP algorithm for building adaptive reduced-order model of nonlinear high-dimensional functions and thus reduce the CPU costs of numerical simulations that rely on these libraries. Several machine learning techniques have been used: encoding based on deep neural networks, community clustering, surrogate modeling and classification in a three-step initial learning phase. Secondly, the definition of an extrapolation metric, followed by a sequential procedure to efficiently allocate more resources and retrain the interpolator network allowed the adaptation of the model to new inputs during real-time usage. 

The algorithm was successfully tested on three direct numerical simulations of hypersonic flows in chemical nonequilibrium. Despite missing information in the initial training, the active learning procedure enhanced the model's versatility and ensured its accuracy even in the presence of evolving flow features with maximum error of the order of 10\%. Moreover, the total time to solution was reduced by up to 70\% when using the original, expensive function. 

This computational framework can be readily ported into other application fields to accelerate simulations that also rely on high-dimensional functions to model complex flow behavior such as combustion, phase-change or fluid-particle interactions.

Finally, future work of the algorithm will consider the active learning of the first two preprocessing steps. For instance, in complex scenarios, different inputs could be projected onto the same location if the off-line training was not sufficient to properly learn the low-dimensional manifold. It might therefore be of interest to adapt the low-dimensional manifold on-the-fly as well. Secondly, if one of the Newman clusters exceedingly grows during the online learning phase, the performance of the corresponding surrogate model would decrease drastically. The cluster could also lose its inner consistency in terms of function dynamics. An interesting approach would be to initialize new Newman clusters online based on a specific criterion.

\section*{Acknowledgment}
This work was supported by the Imperial College London—CNRS PhD Joint Program and was granted access to the HPC/AI resources of TGCC under allocations No. 2021-A0102B12426 and No. 2022-A0122B13432 made by GENCI. Part of the calculations were also performed using MeSU computing platform at Sorbonne University.

\section*{Code availability}
The source code associated with the implementation of RONAALP described in this paper is available on GitHub:

https://github.com/cscherding/RONAALP \newline

Please feel free to explore, use, and provide feedback. If you encounter any issues or have questions, don't hesitate to open an issue on GitHub.



\bibliographystyle{elsarticle-harv} 
\bibliography{Thesis.bib}






\end{document}